# THERMODYNAMICS OF MIXTURES CONTAINING A VERY STRONGLY POLAR COMPOUND. 12. SYSTEMS WITH NITROBENZENE OR 1-NITROALKANE AND HYDROCARBONS OR 1-ALKANOLS

JUAN ANTONIO GONZÁLEZ[(1)*], FERNANDO HEVIA[(1)], LUIS FELIPE SANZ[(1)], ISAÍAS GARCÍA DE LA FUENTE[(1)] AND CRISTINA ALONSO-TRISTÁN[(2)]

[(1)] G.E.T.E.F., Departamento de Física Aplicada, Facultad de Ciencias, Universidad de Valladolid, Paseo de Belén, 7, 47011 Valladolid, Spain,

*e-mail: jagl@termo.uva.es; Tel: +34-983-423757

[(2)] Dpto. Ingeniería Electromecánica. Escuela Politécnica Superior. Avda. Cantabria s/n. 09006 Burgos, (Spain)

**Abstract**

Mixtures involving nitrobenzene and hydrocarbons, or 1-alkanols and 1-nitroalkane, or nitrobenzene have been investigated on the basis of a whole set of thermophysical properties available in the literature. The properties considered are: excess molar functions (enthalpies, entropies, isobaric heat capacities, and volumes), vapour-liquid and liquid-liquid equilibria, permittivities or dynamic viscosities. In addition, the mixtures have been studied by means of the application of the DISQUAC, ERAS, and UNIFAC models, and using the formalism of the concentration-concentration structure factor. The corresponding interaction parameters in the framework of the DISQUAC and ERAS models are reported. In alkane mixtures, dipolar interactions between 1-nitroalkane molecules are weakened when the size of the polar compound increases, accordingly with the relative variation of their effective dipolar moment. Dipolar interactions are stronger in nitrobenzene solutions than in those containing the smaller 1-nitropropane, although both nitroalkanes have very similar effective dipole moment (aromaticity effect). Systems with 1-alkanols are characterized by dipolar interactions between like molecules which sharply increases when the alkanol size increases. Simultaneously, interactions between unlike molecules become weaker, as the OH group is then more sterically hindered. Interactions between unlike molecules are stronger in systems with nitromethane than in nitrobenzene solutions. The replacement of nitromethane by nitroethane in systems with a given 1-alkanol leads to strengthen those effects related with the alcohol self-association. Permittivity data and results on Kirkwood's correlation factors show that the addition of 1-alkanol to a nitroalkane leads to cooperative effects, which increase the dipolar polarization of the solution, in such way that the destruction of the existing structure in pure liquids is partially counterbalanced. This effect is less important when longer 1-alkanols are involved.

Keywords: nitroalkanes; 1-alkanols; thermophysical data; models; dipolar interactions

## 1. Introduction

Nitroalkanes are aprotic solvents of high polarity as it is demonstrated by their large dipole moments (3.56 D (nitromethane); 3.60 D, (nitroethane); 4.0 (nitrobenzene)) [1]. They have many applications. For example, nitromethane is widely used in the manufacture of pharmaceuticals, pesticides or fibers. The industrial interest on the chemistry of nitrobenzene mixtures is due to this compound plays an essential role in the aniline production, and in the preparation of other substances as dyes, paint solvents or the analgesic paracetamol. Unfortunately, it is highly toxic and the hazardous effects to soil, groundwater [2,3] and human health [4,5] must be taken into account.

There is little evidence that nitroalkanes are self-associated in the pure state [6-9]. However, as a consequence of their large $\mu$ values, strong dipolar interactions exist between nitroalkane molecules, and binary mixtures formed by 1-alkanol (from 1-butanol) and nitromethane show liquid-liquid equilibrium (LLE) curves with upper critical solution temperatures (UCST) ranged between 291.1 K (1-butanol) [10,11] and 352.6 K (1-pentadecanol) [11,12]. Similarly, the UCST of the 1-decanol + nitroethane system is 294.1 K [11,13]. In addition, rather large positive values of molar excess Gibbs energies ($G_{\mathrm{m}}^{\mathrm{E}}$), and of enthalpies ($H_{\mathrm{m}}^{\mathrm{E}}$) are encountered for the methanol, or ethanol or 1-propanol, or 1-butanol + nitromethane mixtures [14-16]. That is, 1-alkanol + 1-nitroalkane systems are characterized by positive deviations from the Raoult's law. Interestingly, non-random effects in the mentioned solutions have been investigated by measuring isobaric excess molar heat capacities ($C_{p\mathrm{m}}^{\mathrm{E}}$) [17-20], a very useful magnitude to gain insights into the variation of the solution structure with concentration. In fact, it is well-known that mixtures of the type polar compound + alkane, at temperatures in the vicinity of the critical one, are characterized by W-shaped $C_{p\mathrm{m}}^{\mathrm{E}}$ curves, where non-random effects appear at intermediate compositions [19,21].

Regarding to nitrobenzene systems, a large database exists containing LLE measurements for alkane solutions. Their critical temperatures vary from 291.9 K for the heptane system [22], up to 309.7 K for the hexadecane mixture [23]. These data have been used for the determination of the critical exponents [24,25]. Special attention has been also paid to the dielectric behaviour of these systems near the critical point [26,27].

The main purpose of the present work is to get a deeper understanding of the interactions and structure of nitrobenzene + hydrocarbon mixtures and of 1-alkanol + 1-nitroalkane, or + nitrobenzene systems. At this end, a whole set of experimental data available in the literature, $H_{\mathrm{m}}^{\mathrm{E}}$, $C_{p\mathrm{m}}^{\mathrm{E}}$, excess molar volumes ($V_{\mathrm{m}}^{\mathrm{E}}$), vapour-liquid equilibria (VLE), LLE, permittivities ($\varepsilon_{\mathrm{r}}$), or dynamic viscosities ($\eta$), are analyzed. Of particular interest is the investigation of the aromaticity effect, by means of the study of nitrobenzene solutions. In addition, the selected systems are also treated in the framework of the DISQUAC [28,29] and ERAS [30] models, and the results are compared with those obtained from UNIFAC (Dortmund version) using interaction parameters from the literature [31,32]. The systems are also investigated using the concentration-concentration structure factor ($S_{CC}(0)$) formalism [33], based on the Bhatia-Thorton partial structure factors [34]. The $S_{CC}(0)$ formalism is concerned with the study of fluctuations in the number of molecules regardless of the components, the fluctuations in the mole fraction and the cross fluctuations, and arises from the generalization of the Bhatia-Thorton partial structure factors to link the asymptotic behaviour of the ordering potential to the interchange energy parameters in the semi-phenomenological theories of thermodynamic properties of liquid solutions [35-37]. Thus, we continue our detailed programme concerned with the research of 1-alkanol + strong polar compound mixtures. Within this programme, we have studied mixtures involving, e.g., sulfolane [38], tertiary amides [39,40], or nitriles [41].

## 2. Models

### *2.1 DISQUAC*

The group contribution model DISQUAC is based on the rigid lattice theory developed by Guggenheim [42]. Some of its more relevant features are now briefly summarized. (i) The geometrical parameters of the mixture compounds, total molecular volumes, $r_i$, surfaces, $q_i$, and the molecular surface fractions, $\alpha_{si}$, are calculated additively on the basis of the group volumes $R_G$ and surfaces $Q_G$ recommended by Bondi [43]. At this end, the volume $R_{CH4}$ and surface $Q_{CH4}$ of methane are taken arbitrarily as equal to 1 [44]. The geometrical parameters for the groups used in this work can be found elsewhere [44-48]. (ii) The partition function is factorized into two terms. The excess functions are the result of two contributions: a dispersive (DIS) term arising from the contribution from the dispersive forces; and a quasichemical (QUAC) term which comes from the anisotropy of the field forces created by the solution molecules. For $G_m^E$, a combinatorial term, $G_m^{E,COMB}$, represented by the Flory-Huggins equation [44,47] must be also included. Thus,

$$G_m^E = G_m^{E,DIS} + G_m^{E,QUAC} + G_m^{E,COMB} \quad (1)$$

$$H_m^E = H_m^{E,DIS} + H_m^{E,QUAC} \quad (2)$$

(iii) The interaction parameters change with the molecular structure of the mixture components; (iv) The coordination number is assumed to be the same for all the polar contacts ($z$ = 4). This is a very important shortcoming of the model, and is partially removed via the hypothesis of considering structure dependent interaction parameters.

The equations used to calculate the DIS and QUAC contributions to $G_{\mathrm{m}}^{\mathrm{E}}$ and $H_{\mathrm{m}}^{\mathrm{E}}$ are given elsewhere [29,49]. The temperature dependence of the interaction parameters is expressed in terms of the DIS and QUAC interchange coefficients [29,49], $C_{\mathrm{st},l}^{\mathrm{DIS}};C_{\mathrm{st},l}^{\mathrm{QUAC}}$ where s ≠ t are two contact surfaces present in the mixture and $l$ = 1 (Gibbs energy; $C_{\mathrm{st},1}^{\mathrm{DIS/QUAC}} = g_{\mathrm{st}}^{\mathrm{DIS/QUAC}}(T_{\mathrm{o}})/RT_{\mathrm{o}}$); $l$ = 2 (enthalpy, $C_{\mathrm{st},2}^{\mathrm{DIS/QUAC}} = h_{\mathrm{st}}^{\mathrm{DIS/QUAC}}(T_{\mathrm{o}})/RT_{\mathrm{o}}$)), $l$ = 3 (heat capacity, $C_{\mathrm{st},3}^{\mathrm{DIS/QUAC}} = c_{\mathrm{pst}}^{\mathrm{DIS/QUAC}}(T_{\mathrm{o}})/R$)). $T_{\mathrm{o}}$ = 298.15 K is the scaling temperature and *R*, the gas constant. The equations can be found elsewhere [29,49].

As in previous applications, DISQUAC calculations on LLE were conducted taking into account that the values of the mole fraction $x_1$ of component 1 ($x_1^{'}, x_1^{''}$) relating to the two phases in equilibrium are such that the functions $G_{\mathrm{m}}^{\mathrm{M'}}, G_{m}^{M''}$ ($G_{\mathrm{m}}^{\mathrm{M}} = G_{\mathrm{m}}^{\mathrm{E}} + G_{\mathrm{m}}^{\mathrm{ideal}}$) have a common tangent [50].

*2.2 ERAS*

Some important features of the model are the following. (i) The excess functions are calculated as the sum of two terms. One is linked to hydrogen-bonding effects (the chemical contribution, $X_{\mathrm{m,chem}}^{\mathrm{E}}$), and the other is related to non-polar van der Waals' interactions including free volume effects (physical contribution, $X_{\mathrm{m,phys}}^{\mathrm{E}}$). Equations for $X_{\mathrm{m}}^{\mathrm{E}} = H_{\mathrm{m}}^{\mathrm{E}}$, $V_{\mathrm{m}}^{\mathrm{E}}$ are given elsewhere [49]. (ii) It is assumed that only consecutive linear association occurs. The related chemical equilibrium constant ($K_{\mathrm{A}}$) is independent of the chain length of the associated species (1-alkanols), according to the equation:

$$\mathrm{A_m} + \mathrm{A} \leftrightarrow \mathrm{A_{m+1}} \quad (3)$$

with *m* ranging from 1 to $\infty$. The cross-association between a self-associated species $A_{\mathrm{m}}$ and a non self-associated compound *B* (in the present investigation, 1-nitroalkanes or nitrobenzene) is described by

$$A_{\mathrm{m}} + B \xleftrightarrow{K_{\mathrm{AB}}} A_{\mathrm{m}}B \quad (4)$$

The association constants ($K_{\mathrm{AB}}$) of equation (4) are also assumed to be independent of the chain length. Equations (3) and (4) are characterized by $\Delta h_{\mathrm{i}}^{*}$, the enthalpy of the reaction that corresponds to the hydrogen-bonding energy, and by the volume change ($\Delta v_{\mathrm{i}}^{*}$) related to the formation of the linear chains. (iii) The $X_{\mathrm{m,phys}}^{\mathrm{E}}$ term is derived from the Flory's equation of state [51], which is assumed to be valid not only for pure compounds but also for the mixture [52,53].

$$\frac{\bar{P}_{\mathrm{i}}\bar{V}_{\mathrm{i}}}{\bar{T}_{\mathrm{i}}} = \frac{\bar{V}_{\mathrm{i}}^{1/3}}{\bar{V}_{\mathrm{i}}^{1/3}-1} - \frac{1}{\bar{V}_{\mathrm{i}}\,\bar{T}_{\mathrm{i}}} \qquad (5)$$

where i = A,B or M (mixture). In equation (11), $\bar{V}_{\mathrm{i}} = V_{\mathrm{mi}} / V_{\mathrm{mi}}^{*}$; $\bar{P}_{\mathrm{i}} = P / P_{\mathrm{i}}^{*}$; $\bar{T}_{\mathrm{i}} = T / T_{\mathrm{i}}^{*}$ are the reduced volume, pressure and temperature respectively. The pure component reduction parameters $V_{\mathrm{mi}}^{*}$, $P_{\mathrm{i}}^{*}$, $T_{\mathrm{i}}^{*}$ are determined from *P-V-T* data (density, $\alpha_{\mathrm{p}}$, isobaric thermal expansion coefficient, and isothermal compressibility, $\kappa_{T}$), and association parameters [52,53]. The reduction parameters for the mixture $P_{\mathrm{M}}^{*}$ and $T_{\mathrm{M}}^{*}$ are calculated from mixing rules [52,53]. The total relative molecular volumes and surfaces of the compounds were calculated additively using the Bondi´s method [43].

### *2.3 Modified UNIFAC (Dortmund version)*

This version of UNIFAC [31,32] differs from the original UNIFAC model [54] by the combinatorial term and the temperature dependence of the interaction parameters. The equations used to calculate $G_{\mathrm{m}}^{\mathrm{E}}$ and $H_{\mathrm{m}}^{\mathrm{E}}$ are obtained from the fundamental equation for the activity coefficient $\gamma_{\mathrm{i}}$ of component i:

$$\ln \gamma_{i} = \ln \gamma_{i}^{COMB} + \ln \gamma_{i}^{RES} \qquad (6)$$

where $\ln \gamma_{i}^{COMB}$ and $\ln \gamma_{i}^{RES}$ represent the combinatorial and residual term, respectively. Equations are available elsewhere [49]. In Dortmund UNIFAC, two main groups, OH and $CH_3OH$, are defined for predicting thermodynamic properties of mixtures with alkanols. The main group OH is subdivided in three subgroups: OH(p), OH(s) and OH(t) for the representation of primary, secondary and tertiary alkanols, respectively. The $CH_3OH$ group is a specific group for methanol solutions. In the case of nitroalkanes and nitrobenzene two main groups exist. The main group CNO2 is divided in three subgroups: $CH_3NO_2$ for nitromethane; $CH_2NO_2$ for the remainder 1-nitroalkanes, and $CHNO_2$ for 2-nitroalkanes. There is also a main group, $ACNO_2$, for nitrobenzene. The subgroups within the same main group have different geometrical parameters, and identical group energy-interaction parameters. It is remarkable that the geometrical parameters, the relative van der Waals volumes and the relative van der Waals surfaces are not calculated from molecular parameters like in the original UNIFAC, but fitted together with the interaction parameters to the experimental values of the thermodynamic properties considered. The geometrical and interaction parameters were taken from literature and used without modifications [32]. No interaction parameters are available for the methanol + nitrobenzene system.

*2.4 The concentration-concentration structure factor*

Mixture structure can be investigated by means of the $S_{CC}(0)$ function [33,35,36,55,56]:

$$S_{CC}(0) = \frac{RT}{(\partial^2 G^M / \partial x_1^2)_{P,T}} = \frac{x_1 x_2}{D} \quad (7)$$

with

$$D = \frac{x_1 x_2}{RT}(\partial^2 G^M / \partial x_1^2)_{P,T} = 1 + \frac{x_1 x_2}{RT}\left(\frac{\partial^2 G_m^E}{\partial x_1^2}\right)_{P,T} \quad (8)$$

$D$ is a function closely related to thermodynamic stability [57-59]. For ideal mixtures, $G_m^{E,id} = 0$ (excess Gibbs energy of the ideal mixture); $D^{id} = 1$ and $S_{CC}(0) = x_1 x_2$. From stability conditions, $S_{CC}(0) > 0$. If a system is close to phase separation, $S_{CC}(0)$ must be large and positive ($\infty$, if the mixture presents a miscibility gap). If compound formation between components exists, $S_{CC}(0)$ must be very low (0, in the limit). Therefore, $S_{CC}(0) > x_1 x_2$ ($D < 1$) indicates that the dominant trend in the system is the homocoordination (separation of the components). The mixture is then less stable than the ideal. If $0 < S_{CC}(0) < x_1 x_2 = S_{CC}(0)^{id}$, ($D > 1$), the fluctuations in the system have been removed, and the main feature of the solution is compound formation (heterocoordination). The system is then more stable than ideal. In summary, $S_{CC}(0)$ is an useful magnitude to evaluate the non-randomness in the mixture [55,56]. In this work, we have used DISQUAC to evaluate $S_{CC}(0)$ for a number of mixtures.

## 3. Adjustment of model parameters

*3.1 DISQUAC interaction parameters*

In terms of DISQUAC, the studied systems are regarded as possessing the following types of surfaces: (i) type a, aliphatic ($CH_3$, $CH_2$, in *n*-alkanes, or toluene, or 1-nitroalkane, or 1-alkanols); (ii) type r ($NO_2$ in 1-nitroalkanes or nitrobenzenee); (iii) type s (s = b, $C_6H_6$, or $C_6H_5$ in benzene, toluene or nitrobenzene; s = c-$CH_2$ in cyclohexane; s = h, OH in 1-alkanols).

The general procedure applied in the estimation of the interaction parameters have been explained in detail in earlier works [29,49]. Final values of our fitted parameters are listed in Tables 1 and 2. Some important remarks are provided below.

*3.1.1 Nitrobenzene + benzene*

This system is only characterized by the (b,r) contact, which is assumed to be represented by DIS interaction parameters. Such choice is supported by the low experimental $H_{\mathrm{m}}^{\mathrm{E}}$ values of this system (261 J·mol$^{-1}$ at equimolar composition and 293.15 K [60]). The $C_{\mathrm{br},1}^{\mathrm{DIS}}$ coefficient was obtained from data on activity coefficients at infinite dilution [61]. Final parameters are given in Table 1.

*3.1.2 Nitrobenzene + alkane, or + toluene*

Mixtures with alkane are built by three contacts: (a,b), (b,r) and (a,r). The interaction parameters for the (a,b) contacts are dispersive and are known from the research of alkyl-benzene + alkane systems [45]. The interaction parameters of the (b,r) contacts are already known and thus only those corresponding to the (a,r) contacts must be determined (Table 1). As in other many applications, we have used $C_{\mathrm{ar},l}^{\mathrm{QUAC}} = C_{\mathrm{cr},l}^{\mathrm{QUAC}}$ ($l$ = 1,2,3). That is, the QUAC coefficients for the (a,r) and (c,r) contacts are independent of the alkane [29,40,62]. In addition, the $C_{\mathrm{ar},1}^{\mathrm{DIS}}$ coefficient is assumed to be dependent on the chain length of the *n*-alkane in order to get improved results for the coordinates of the critical points (see below).

The system involving toluene is characterized by the same contacts. Here, we have held the interaction parameters for the (a,b) and (a,r) contacts and determined newly those for the (b,r) contact (Table 1), assuming that it is dispersive.

*3.1.3 1-Alkanol + 1-nitroalkane*

The contacts present in these solutions are: (a,h), (a,r) and (h,r). The interaction parameters for the (a,h) contacts are described by DIS and QUAC interaction parameters, previously determined from the study of 1-alkanol + *n*-alkane systems [46,63-65], and the $C_{\mathrm{ar},l}^{\mathrm{DIS/QUAC}}$ (l = 1,2,3) coefficients are also known from the corresponding treatment of 1-nitroalkane + alkane systems [48]. Therefore, only the interaction parameters for the (h,r) have to be obtained (Table 2). Detailed calculations showed that the QUAC coefficients could be assumed to be independent of the 1-alkanol (from 1-propanol). Consequently, these parameters were fixed and the DIS ones determined. We remark that, for a similar reason to that given for nitrobenzene + *n*-alkane systems, the $C_{\mathrm{hr},1}^{\mathrm{DIS}}$ coefficient is considered to be dependent on the alkanol size.

*3.1.4 1-Alkanols + nitrobenzene*

We have now six contacts: (a,b), (a,h), (a,r), (b,h), (b,r) and (h,r). The contacts (b,h) are represented by DIS and QUAC interaction parameters, which are already known from the study of 1-alkanol + toluene mixtures [65,66]. Thus, only the interaction parameters for the (h,r) contacts have to be determined (Table 2) as those for the remainder contacts are known. The procedure is similar to that just explained.

*3.2 Adjustment of ERAS parameters*

Values of $V_{\mathrm{mi}}, V_{\mathrm{mi}}^{*}$ and $P_{\mathrm{i}}^{*}$ of pure compounds at $T$ = 298.15 K, needed for calculations, have been taken from the literature in the case of 1-alkanols [67], and are listed in Table S1 of supplementary material for 1-nitroalkanes or nitrobenzene. For the 1-alkanols, $K_{\mathrm{A}}$, $\Delta h_{\mathrm{A}}^{*}$ (= – 25.1 kJ·mol$^{-1}$) and $\Delta v_{\mathrm{A}}^{*}$ (= – 5.6 cm$^{3}$·mol$^{-1}$) are known from $H_{\mathrm{m}}^{\mathrm{E}}$ and $V_{\mathrm{m}}^{\mathrm{E}}$ data for the corresponding mixtures with alkanes. These values have been used in many other applications [67]. The binary parameters to be fitted against $H_{\mathrm{m}}^{\mathrm{E}}$ and $V_{\mathrm{m}}^{\mathrm{E}}$ data available in the literature for 1-alkanol + 1-nitroalkane or + nitrobenzene systems are then $K_{\mathrm{AB}}$, $\Delta h_{\mathrm{AB}}^{*}$, $\Delta v_{\mathrm{AB}}^{*}$ and $X_{\mathrm{AB}}$. They are collected in Table 3.

## 4. Theoretical results

Results from DISQUAC on phase equilibria, $H_{\mathrm{m}}^{\mathrm{E}}$ and $C_{\mathrm{pm}}^{\mathrm{E}}$ are shown in Tables 4-8 and in Figures 1-7 (see also Figure S1 of supplementary material). Tables 4 and 7 contain relative deviations for pressure and $H_{\mathrm{m}}^{\mathrm{E}}$ , respectively, defined as

$$\sigma_{\mathrm{r}}(P) = \{\frac{1}{N}\sum\left[\frac{P_{\mathrm{exp}} - P_{\mathrm{calc}}}{P_{\mathrm{exp}}}\right]^{2}\}^{1/2} \qquad (9)$$

$$dev(H_{\mathrm{m}}^{\mathrm{E}}) = \{\frac{1}{N}\sum\left[\frac{H_{\mathrm{m,exp}}^{\mathrm{E}} - H_{\mathrm{m,calc}}^{\mathrm{E}}}{H_{\mathrm{m,exp}}^{\mathrm{E}}(x_1 = 0.5)}\right]^{2}\}^{1/2} \qquad (10)$$

where, $N$ stands for the number of data points. ERAS results on $H_{\mathrm{m}}^{\mathrm{E}}$ for 1-alkanol + nitromethane or + nitrobenzene systems are shown in Table 7 (Figures 5 and 6). Some ERAS calculations on $V_{\mathrm{m}}^{\mathrm{E}}$ are collected in Table S2 (see Figure S2 of supplementary material). Figure S1 compares, as an example, ERAS results with experimental $G_{\mathrm{m}}^{\mathrm{E}}$ values for the methanol + nitromethane mixture. Results from the application of the UNIFAC model on VLE, $H_{\mathrm{m}}^{\mathrm{E}}$ and $C_{\mathrm{pm}}^{\mathrm{E}}$ are collected in Tables 4, 7 and 8.

## 5. Discussion

Hereafter, we are referring to values of the thermodynamic excess functions at 298.15 K and equimolar composition. The number of C atoms in 1-alkanols and in 1-nitroalkanes are represented by $n_{\mathrm{OH}}$ and $n_{\mathrm{NO2}}$, respectively. The impact of polarity on bulk properties can be examined through the effective dipole moment, $\bar{\mu}$ , defined by [49,57,68,69]:

$$\bar{\mu} = \left[ \frac{\mu^2 N_A}{4\pi\varepsilon_0 V_{\mathrm{m}} k_B T} \right]^{1/2} \tag{11}$$

Here, $\mu$ is the dipole moment; $N_A$ the Avogadro's number, $\varepsilon_0$ the permittivity of the vacuum, $V_{\mathrm{m}}$ the molar volume, and $k_B$ the Boltzmann's constant. Values of $\bar{\mu}$ for 1-nitroalkanes and nitrobenzene are collected in Table 9. As for a pure polar liquid, the potential energy related to dipole-dipole interactions is, in first approximation, proportional to $(-\bar{\mu}^4/r^6)$ [70] or more roughly to $(-\bar{\mu}^4/V_{\mathrm{m}}^2)$ [71] ($r$ is the distance between dipoles), dipolar interactions between nitroalkane molecules decrease in the order: nitromethane ($\bar{\mu}$ = 1.855) > nitroethane ($\bar{\mu}$ = 1.625) > 1-nitropropane ($\bar{\mu}$ = 1.453). For nitrobenzene, $\bar{\mu}$ = 1.510. Dipolar interactions between 1-alkanol molecules are much weaker and also decrease with the increasing of molecular size: $\bar{\mu}$ = 1.023 ($n_{\mathrm{OH}}$ = 1) > 0.852 ($n_{\mathrm{OH}}$ = 2) > 0.752 ($n_{\mathrm{OH}}$ = 3) > 0.664 ($n_{\mathrm{OH}}$ = 4) > 0.580 ($n_{\mathrm{OH}}$ = 6) [49].

### *5.1 Nitrobenzene + hydrocarbon mixtures*

Firstly, we must remark that nitromethane or nitroethane + alkane systems show LLE curves characterized by UCSTs, which in the case of nitromethane solutions are very high: 387.2 K and 398.1 K for the systems with octane [72] and decane [73], respectively. The critical temperatures of nitroethane + octane (314.5 K), or + decane (325.8 K) [74] are lower and $H_{\mathrm{m}}^{\mathrm{E}}$/J·mol$^{-1}$ values of heptane mixtures decrease when $n_{\mathrm{NO2}}$ is increased: 1593 ($n_{\mathrm{NO2}}$ = 3); 1390 ($n_{\mathrm{NO2}}$ = 4); 1220 ($n_{\mathrm{NO2}}$ = 5) [48]. Similarly, $H_{\mathrm{m}}^{\mathrm{E}}$($C_6H_{12}$)/J·mol$^{-1}$ = 1690 ($n_{\mathrm{NO2}}$ = 2); 1549 ($n_{\mathrm{NO2}}$ = 3); 1370 ($n_{\mathrm{NO2}}$ = 4); 1225 ($n_{\mathrm{NO2}}$ = 5) [48]. These experimental results are in agreement with the existence of strong dipolar interactions between nitroalkane molecules and reveal that the mentioned interactions become weaker when $n_{\mathrm{NO2}}$ increases, accordingly with the relative variation of $\bar{\mu}$. Regarding nitrobenzene + alkane systems, UCST/K = 293.01 K (octane) [75]; 295.96 K (decane) [76] and $H_{\mathrm{m}}^{\mathrm{E}}$(cyclohexane, $T$ = 293.15 K) = 1654 J·mol$^{-1}$ [77]; 1447 (hexane) [78]. Such set of data suggests that dipolar interactions between nitrobenzene molecules are stronger than those between 1-nitropropane molecules, even when the distance between nitrobenzene molecules is larger (Table 9). This can be ascribed to the existence of intramolecular effects between the phenyl and the $NO_2$ groups of nitrobenzene which lead to enhanced dipolar interactions. Intermolecular effects between the phenyl ring and the polar group of a given aromatic polar molecule is encountered in many systems. Thus, 1-alkanol (from ethanol) [46,79] or 1-alkylamine (from ethylamine) [80] + heptane mixtures are miscible at any composition at 298.15 K, while the UCST of the corresponding solutions with phenol or aniline are 327.3 K [81] and 343.1 K [82], respectively. Similarly, UCST(decane)/K = 266.8 (2-propanone) [83], 277.4 (acetophenone) [84], or UCST(dodecane)/K = 284.7 (butanenitrile) [85]; 293.1 (benzonitrile) [86].

It is interesting to conduct a short comparison between mixtures containing 1-nitroalkanes and alkanes or benzene. For example, $H_m^E$ (benzene)/J·mol$^{-1}$ = 790 ($n_{NO2}$ = 1); 63 ($n_{NO2}$ = 3) [87]; and 261 (nitrobenzene, $T$/K = 293.15) [60]. That is, when the *n*-alkane is replaced by benzene, mixtures become miscible and $H_m^E$ values decrease, which can be ascribed to the new interactions between unlike molecules created upon mixing (intermolecular effects). This is a rather general trend, as the following values reveal: $H_m^E$($C_6H_6$)/ J·mol$^{-1}$ = 751 (aniline) [88]; 860 (phenol, $T$ = 313.15 K) [89]; 125 (acetophenone) [90]; 138 (2-propanone) [91]; – 66 (butanenitrile) [92]; 32 (benzonitrile) [93]. 1-Alkanols are a remarkable exception and $H_m^E$(1-alkanol ($\neq$ methanol) + heptane) > $H_m^E$(1-alkanol + benzene). Thus for 1-hexanol mixtures, $H_m^E$/J·mol$^{-1}$ = 527 (heptane) [94]; 1141 (benzene) [95]. Aromatic hydrocarbons are better breakers of the alcohol self-association than *n*-alkanes.

*5.2 1-alkanol + 1-nitroalkane*

We start remarking that the existence, for these solutions, of LLE curves with relatively high UCST values (see Introduction Section) show that dipolar interactions are very important. For the sake of comparison, we provide UCST values of hexane mixtures with tertiary amides together with dipole moments of these very polar substances. Thus, UCST/K = 337.7 [96] (*N,N*-dimethylformamide; $\mu$ = 3.68 D [1]); 305.3 [97] (*N,N*-dimethylacetamide; $\mu$ = 3.71 D [1]); 324.6 [98] (*N*-methylpyrrolidone; $\mu$ = 4.09 D [99]. For the 1-hexanol + nitromethane system, UCST = 308.7 K [100]. This value is much lower than the corresponding result for the nitromethane + hexane mixture (375.4 K [72]). That is, the replacement of an *n*-alkane by an isomeric 1-alkanol leads to a decreased UCST value, which reveals that 1-alkanols are better breakers of the dipolar interactions between nitroalkane molecules. The existence of alkanol-nitroalkane interactions is supported by the fact that systems with $n_{OH}$ = 1-4 and $n_{NO2}$ = 1 are miscible at 298.15 K and any concentration (see below).

Next, we are going to evaluate the enthalpy of the H-bonds between 1-alkanols and nitroalkanes (termed as $\Delta H_{OH\text{-}NO2}$). Neglecting structural effects [57,101], $H_m^E$ can be considered as the result of three contributions. The positive ones, $\Delta H_{\text{OH-OH}}$, $\Delta H_{\text{NO2-NO2}}$, come, respectively, from the breaking of alkanol-alkanol and nitroalkane-nitroalkane interactions along the mixing process. The negative contribution, $\Delta H_{\text{OH-NO2}}$, is due to the new OH---NO2 interactions created upon mixing. That is [67,102-104]:

$$H_{\text{m}}^{\text{E}} = \Delta H_{\text{OH-OH}} + \Delta H_{\text{NO2-NO2}} + \Delta H_{\text{OH-NO2}} \qquad (12)$$

An evaluation of $\Delta H_{\text{OH-NO2}}$ can be conducted extending the equation (12) to $x_1 \to 0$ [67, 105,106]. Then, $\Delta H_{\text{OH-OH}}$ and $\Delta H_{\text{NO2-NO2}}$ can be replaced by $H_{\text{m1}}^{\text{E},\infty}$ (partial excess molar enthalpy at infinite dilution of the first component) of 1-alkanol or nitroalkane + heptane systems. Thus,

$$\Delta H_{\text{OH-NO2}} = H_{\text{m1}}^{\text{E},\infty}(1-\text{alkanol + nitroalkane})$$

$$-H_{\text{m1}}^{\text{E},\infty}(1-\text{alkanol + heptane}) - H_{\text{m1}}^{\text{E},\infty}(\text{nitroalkane + heptane}) \qquad (13)$$

There are some shortcomings for this estimation of $\Delta H_{\text{OH-NO2}}$ values. (i) Some $H_{\text{m1}}^{\text{E},\infty}$ data used were calculated from $H_{\text{m}}^{\text{E}}$ measurements over the entire mole fraction range. (ii) For 1-alkanol + *n*-alkane systems, it was assumed that $H_{\text{m1}}^{\text{E},\infty}$ is independent of the alcohol, a common approach when applying association theories [30,107-109]. We have used in this work, as in previous applications [67,110], $H_{\text{m1}}^{\text{E},\infty}$ = 23.2 kJ·mol$^{-1}$ [111-113]. Nevertheless, it should be remarked that the values of $\Delta H_{\text{OH-NO2}}$ collected in Table 10 are still meaningful as they were obtained following the same procedure that in other previous investigations, which allows to compare enthalpies of interaction between 1-alkanols and different organic solvents. Inspection of Table 10 shows that $\Delta H_{\text{OH-NO2}}$ increases more or less smoothly with $n_{\text{OH}}$, that is, interactions between unlike molecules become weaker, which may be ascribed to the OH group is more sterically hindered in longer 1-alkanols. On the other hand, the increased UCST values for nitromethane mixtures with longer 1-alkanols ($n_{\text{OH}} \geq 6$) suggests a sharp decrease of the number of interactions between unlike molecules, while interactions between like molecules become strongly dominant.

### 5.2.1 *Molar excess enthalpies and entropies*

The 1-alkanol + nitromethane mixtures are characterized by: (i) large and positive $H_m^E$ values ($H_m^E$/J·mol$^{-1}$ = 1265 ($n_{OH}$ = 1); 1633 ($n_{OH}$ = 2); 1911 ($n_{OH}$ = 3); 2131 ($n_{OH}$ = 4) [16]; 2781 ($n_{OH}$ = 6), $T$ = 313.15 K [114]). (ii) Symmetrical $H_m^E$ curves, which, at the middle of the concentration range, are more or less flattened as consequence of the proximity of the UCST (Figure 5). (iii) Positive $TS_m^E$ ($= H_m^E - G_m^E$) values: 228 ($n_{OH}$ = 1); 463 ($n_{OH}$ = 2); 1459 ($n_{OH}$ = 6, $T$ = 313.15 K) (values calculated using $G_m^E$/J·mol$^{-1}$ = 1040 ($n_{OH}$ = 1); 1170 ($n_{OH}$ = 2) [14]; 1322 ($n_{OH}$ = 6, DISQUAC value at $T$ = 313.15 K)). These features support our previous conclusion on the relevance of dipolar interactions in the present systems. Note that several typical properties of 1-alkanol + alkane mixtures are: (i) low $H_m^E$ values, as alkanes are poor breakers of the alkanols self-association; (ii) for the same reason, the $H_m^E$ curves are skewed to low alcohol mole fractions; (iii) very negative $TS_m^E$ values. For example, for the 1-propanol + hexane system, $G_m^E$ = 1295 [115], $H_m^E$ = 533 [116] and $TS_m^E = -762$ (all values in J·mol$^{-1}$).

The $H_m^E(n_{OH})$ variation of 1-alkanol + nitromethane systems can be explained as follows. (i) Interactions between unlike molecules are weakened (lower $|\Delta H_{OH\text{-}NO2}|$ values) when $n_{OH}$ increases. (ii) Dipolar interactions between like molecules become more important at this condition, as the $TS_m^E(n_{OH})$ variation shows. Note the very high result for $TS_m^E$ of the 1-hexanol solution at 313.15 K, a temperature only 4.45 K higher than the UCST. Dipolar interactions are weakened when nitromethane is replaced by nitroethane, as the UCST values of the corresponding 1-nonanol mixtures reveal: (328 K ($n_{NO2}$ =1) [11,117]; 283.6 K ($n_{NO2}$ = 2) [11,13].

### 5.2.3 *Molar excess volumes*

It is well stated that $V_m^E$ is the result of different contributions. Those which are positive arise from the breaking of interactions between like molecules. Interactions between unlike molecules and structural effects (changes in free volume, differences in size and shape between the system components, interstitial accommodation) contribute negatively to $V_m^E$. Thus, the positive $V_m^E$/(cm$^3$·mol$^{-1}$) values of the systems 1-nitropropane + $C_6H_{12}$ (0.690) [118] or 1-propanol + heptane (0.271) [119] indicate that the main contribution to $V_m^E$ comes from the disruption of the dipolar interactions between 1-nitropropane molecules and from the breaking of the alcohol self-association, respectively. The lower $V_m^E$/(cm$^3$·mol$^{-1}$) value of 1-propanol +

nitromethane mixture (0.236 [18]) newly points out to the existence of interactions between unlike molecules in this kind of systems, which are also characterized by strong structural effects. In fact, $H_m^E$ and $V_m^E$ values of some solutions are of opposite sign (e.g., $V_m^E$ (methanol + nitromethane) = – 0.152 cm$^3$·mol$^{-1}$ [17]). This is a typical feature of mixtures where strong structural effects exist [120,121]. In addition, the positive $V_m^E$ values encountered for these solutions (0.342 cm$^3$·mol$^{-1}$ for 1-butanol + nitromethane [19], see also above) are rather low, if one takes into account the very large $H_m^E$ values involved. On the other hand, for a given 1-nitroalkane, both $H_m^E$ and $V_m^E$ increase in line with $n_{OH}$. That is, the $V_m^E$ ($n_{OH}$) variation is closely related to that of the corresponding interactional contribution to this excess function. Interestingly, $V_m^E$ is lower for the methanol + nitromethane system (– 0.168 cm$^3$·mol$^{-1}$ [122] than for the corresponding mixture with nitroethane (– 0.141 cm$^3$·mol$^{-1}$ [123], values at 293.15 K). An inversion of this behaviour is observed for solutions with $n_{OH} \geq 2$. Thus, $V_m^E$ (1-propanol)/cm$^3$·mol$^{-1}$: 0.213, ($n_{NO2}$ = 1) [122]; 0.111 ($n_{NO2}$ = 2) [123]. This suggests that effects linked to interactions between unlike molecules are more important in the methanol + nitromethane system, and that those related to the breaking of dipolar interactions between nitroalkane molecules are predominant in systems with $n_{OH} \geq 2$ and $n_{NO2}$ = 1. At the latter conditions, effects due to alcohol self-association become more relevant in solutions with $n_{NO2}$ = 2. Accordingly with this interpretation, the $V_m^E$ curve of the 1-propanol + nitroethane systems is skewed to lower concentrations in the alcohol, while the corresponding curve for 1-propanol + nitromethane is nearly symmetrical [122,123].

#### *5.2.4 Molar excess heat capacities at constant pressure*

It is important to keep in mind that $C_{p,m}^E$ values of 1-alkanol + alkane mixtures are high and positive (11.7 J·mol$^{-1}$·K$^{-1}$ for ethanol + heptane [124]). In addition, $C_{p,m}^E$ increases with the temperature and, at enough high values of this magnitude, decreases [46,125]. For example, in the case of the 1-decanol + decane mixture, $C_{p,m}^E$ ($x_1$ = 0.3925)/J·mol$^{-1}$·K$^{-1}$ = 14.81 (298.15 K); 25.04 (348.15 K); 24.31 (368.15 K) [125]. Mixtures characterized by strong dipolar interactions show low positive $C_{p,m}^E$ values. Thus, $C_{p,m}^E$ /J·mol$^{-1}$·K$^{-1}$ = 4.4 (1-propanol + 2,5,8-trioxanonane) [126]; 0.96 (ethanol + DMF) [127]. Therefore, the $C_{p,m}^E$ value of the methanol + nitromethane mixture (9.3 J·mol$^{-1}$·K$^{-1}$ [17]) reveals that association/solvation effects are still important in this solution. The available $C_{p,m}^E$ data for 1-propanol or 1-butanol + nitromethane mixtures are very large (16.6 [18] and 20.6 [19] J·mol$^{-1}$·K$^{-1}$, respectively), and decrease when the temperature is increased [18,19] (Table 8). These are typical features encountered in systems at temperatures not far from their UCST [21]. In this framework, the observed $C_{\text{p,m}}^{\text{E}}(n_{\text{OH}})$ variation at 298.15 K in nitromethane solutions (Table 8) can be ascribed to an increase of non-random effects related to the proximity of the critical temperature. The $C_{\text{p,m}}^{\text{E}}$ result of the ethanol + 1-nitropropane mixture (14.9 $\text{J·mol}^{-1}\text{·K}^{-1}$ [20]) seems to be higher than the corresponding value of the nitroethane system, which can be explained considering that dipolar interactions are much less relevant in 1-nitropropane mixtures where effects related to alcohol self-association become more relevant.

*5.3 1-alkanol + nitrobenzene*

The $H_{\text{m}}^{\text{E}}$ values of these systems are also large and positive: $H_{\text{m}}^{\text{E}}$ / $\text{J·mol}^{-1}$ = 1109 ($n_{\text{OH}}$ = 1) [128]; 1430 ($n_{\text{OH}}$ = 2); 1946 ($n_{\text{OH}}$ = 4) [129]. Consequently, the main contribution to $H_{\text{m}}^{\text{E}}$ arises from the breaking of interactions between like molecules. The $H_{\text{m}}^{\text{E}}(n_{\text{OH}})$ variation can be explained as above. i.e., in terms of a weakening of the alkanol-nitrobenzene interactions produced when $n_{\text{OH}}$ increases (Table 10), together with the corresponding increase of dipolar interactions between like molecules. The excess molar volumes are negative: −0.191 ($n_{\text{OH}}$ = 3), −0.117 ($n_{\text{OH}}$ = 4), −0.075 ($n_{\text{OH}}$ = 5) ($T$ = 303.15 K) [130], i.e., the contribution to $V_{\text{m}}^{\text{E}}$ from structural effects is here dominant.

On the other hand, comparison of $\Delta H_{\text{OH-NO2}}$ values for systems with nitromethane or nitrobenzene (Table 10) suggests that interactions between unlike molecules are weaker in the latter solutions. However, the comparison between $H_{\text{m}}^{\text{E}}$ values pertaining to different homologous series should be conducted with caution in order to state reliable conclusions. It is known that $H_{\text{m}}^{\text{E}}$ is not only determined by interactional effects, but also by structural effects. In fact, it is more appropriated compare $U_{\text{Vm}}^{\text{E}}$ results (isochoric molar excess internal energy), which can be obtained from [57,101]:

$$U_{\text{Vm}}^{\text{E}} = H_{\text{m}}^{\text{E}} - \frac{\alpha_{\text{p}}}{\kappa_{\text{T}}} T V_{\text{m}}^{\text{E}} \qquad (14)$$

where $\frac{\alpha_p}{\kappa_T} T V_m^{\text{E}}$ represents the equation of state (eos) contribution to $H_{\text{m}}^{\text{E}}$, $\alpha_{\text{p}}$ and $\kappa_T$ are, respectively, the isobaric thermal expansion coefficient and the isothermal compressibility of the mixture. The determination of the eos contribution needs of accurate volumetric data. Here, the very different $V_{\text{m}}^{\text{E}}$ values of 1-alkanol + nitromethane, or + nitrobenzene systems (see above) suggest that the eos term may be decisive when comparing $H_{\rm m}^{\rm E}$ values pertaining to these series. We have roughly determined $U_{\rm Vm}^{\rm E}$ /J·mol$^{-1}$ values for 1-propanol + nitromethane (1826), or + nitrobenzene (1882) systems, and for 1-butanol + nitromethane (2008), or + nitrobenzene (1992) mixtures, assuming the ideal behavior to calculate $\alpha_{\rm p}$ and $\kappa_T$ values of the mixtures. The rather similar results obtained for solutions with a given 1-alkanol point out that the different $H_{\rm m}^{\rm E}$ values are largely due to the different structural effects in the considered systems.

*5.4 Dielectric constants and Kirkwood's correlation factor*

Here, we analyze the permittivity data, $\varepsilon_{\rm r}$, for 1-alkanol + nitromethane, or + nitrobenzene mixtures. The excess permittivities, $\varepsilon_{\rm r}^{\rm E}$, are defined as:

$$\varepsilon_{\rm r}^{\rm E} = \varepsilon_{\rm r} - \phi_1\varepsilon_{\rm r1} - \phi_2\varepsilon_{\rm r2} \qquad (16)$$

where $\phi_{\rm i} = x_{\rm i}V_{\rm mi} / \sum x_{\rm i}V_{\rm mi}$ is the volume fraction. The $\varepsilon_{\rm r}^{\rm E}$ values referred to below are, except when indicated, at 298.15 K and $\phi_{\rm i} = 0.5$. For the present systems $\varepsilon_{\rm r}^{\rm E}$ values are very negative. Thus, $\varepsilon_{\rm r}^{\rm E}$(nitromethane; 293.15 K) = – 1.74 ($n_{\rm OH}$ = 1); – 2.55 ($n_{\rm OH}$ = 2); – 3.20 ($n_{\rm OH}$ = 3) [131,132]; and $\varepsilon_{\rm r}^{\rm E}$(nitrobenzene) = – 2.46 ($n_{\rm OH}$ = 2); – 4.01 ($n_{\rm OH}$ = 3); – 4.23 ($n_{\rm OH}$ = 4) [133]; – 2.67 ($n_{\rm OH}$ = 5); – 0.90 ($n_{\rm OH}$ = 7) ($T$ = 293.15 K) [132,134]. These negative experimental results indicate that the predominant trend in the solutions is the breaking of the alcohol self-association, as well as the disruption of the dipolar interactions between nitroalkane molecules, as such effects lead to a decrease of the dipolar polarization of the mixture [135-137]. The contribution to $\varepsilon_{\rm r}^{\rm E}$ due to interactions between unlike molecules may be either positive or negative, and that depends on the chemical nature of the mixture compounds and of the size and shape of the multimers formed upon mixing by the two components. Negative contributions are encountered when the mentioned multimers form, eg., cyclic structures and have less effective dipole moments than those of the multimers built by the pure components [135]. Interestingly, the $\varepsilon_{\rm r}^{\rm E}$ ($\phi_1$ = 0.5) value of the nitrobenzene + benzene mixture is also very negative (– 4.88 [138]). It is clear that benzene is a good breaker of the dipolar interactions between nitrobenzene molecules. The variation of $\varepsilon_{\rm r}^{\rm E}$(nitrobenzene) with $n_{\rm OH}$ seems to be a rather general trend, as many mixtures behave similarly. For example, $\varepsilon_{\rm r}^{\rm E}$(hexane)= – 1.12 ($n_{\rm OH}$ = 4) [139]; – 2.43 ($n_{\rm OH}$ = 5); – 2.52 ($n_{\rm OH}$ = 7) [140]; – 1.62 ($n_{\rm OH}$ = 10) [141]; or $\varepsilon_{\rm r}^{\rm E}$(cyclohexylamine) = 2.22 ($n_{\rm OH}$ = 1) [142] – 0.27 ($n_{\rm OH}$ = 3); – 0.85 ($n_{\rm OH}$ = 4); – 0.91 ($n_{\rm OH}$ = 7): – 0.41 ($n_{\rm OH}$ = 10) [135]. This behaviour has been explained in terms of the lower and weaker self-association of longer 1-alkanols [135].

We have determined the Kirkwood's correlation factor, $g_K$ , of systems containing 1-alkanols by means of the equation [143-145]:

$$g_K = \frac{9k_B T V_m \varepsilon_0 (\varepsilon_r - \varepsilon_r^\infty)(2\varepsilon_r + \varepsilon_r^\infty)}{N_A \mu^2 \varepsilon_r (\varepsilon_r^\infty + 2)^2} \quad (17)$$

where the symbols have the usual meaning [135]. Details of the calculation procedure have been given previously [135,146]. The equation needed for the determination of $g_K$ for the nitrobenzene + benzene system is different as benzene is a non-polar compound and can be found elsewhere [147]. Physical properties of pure compounds and density data needed for calculations were taken from the literature [1,131,132,148-152]. In absence of experimental measurements on refractive indices, this magnitude was considered as ideal [153]. For 1-alkanol + nitrobenzene systems, results plotted in Figure 8 show that the mixture structure increases smoothly over a rather wide concentration range, and that this increase becomes even softer when $n_{OH}$ increases. Thus, the addition of 1-alkanol to nitroalkane leads to cooperative effects, which increase the total effective dipole moment of the solution, and that partially compensate the destruction of the existing structure in pure liquids. This effect becomes less relevant when longer 1-alkanols are involved. The same behaviour is encountered for 1-alkanol + nitromethane mixtures. Here, it is important to pay attention to the nitrobenzene + benzene system. Figure 8 clearly shows that there is a large loss of structure upon mixing for this mixture, which might be indicative of the existence of random effects

*5.5 Dynamic viscosities*

The discussion is now conducted in terms of deviations of viscosity from the linear behaviour, defined as

$$\Delta\eta = \eta - x_1\eta_1 + x_2\eta_2 \quad (18)$$

Results given below are at equimolar composition. Firstly, we must remark that $\Delta\eta$ values of the studied systems are negative. At 303.15 K, $\Delta\eta$ ($n_{NO2}$ = 1)/mPa·s = – 0.084 ($n_{OH}$ =1); – 0.211 ($n_{OH}$ = 2); – 0.408 ($n_{OH}$ = 3) [122]; $\Delta\eta$ ($n_{NO2}$ = 2)/mPa·s = – 0.066 ($n_{OH}$ =1); – 0.193 ($n_{OH}$ = 2); – 0.407 ($n_{OH}$ = 3) [123]; and $\Delta\eta$ (nitrobenzene)/mPa·s = – 0.043 ($n_{OH}$ =1); – 0.172 ($n_{OH}$ = 2); – 0.342 ($n_{OH}$ = 3); – 0.494 ($n_{OH}$ = 4) [151]. These results can be explained in terms of a higher fluidization of the solution due to the breaking of alcohol self-association and of dipolar interactions between nitroalkane molecules. As viscosity is strongly dependent on the size and shape of the mixture components, we compare now the results of nitrobenzene solutions with those of 1-alkanol + toluene systems at 303.15 K. The latter mixtures show lower values. Thus, $\Delta\eta$ /mPa·s = − 0.056 ($n_{\mathrm{OH}}$ = 1); − 0.228 ($n_{\mathrm{OH}}$ = 2) [154]; − 0.381 ($n_{\mathrm{OH}}$ = 3); − 0.557 ($n_{\mathrm{OH}}$ = 4) [155]. From this comparison, it is possible to conclude that alkanol-nitrobenzene interactions lead to a lower fluidization of the corresponding mixture (higher $\Delta\eta$ values). Of course, interactions between unlike molecules are also present in systems with nitromethane or nitroethane. This can be demonstrated by comparing ($\Delta\eta$ /mPa·s) data for mixtures with $n_{\mathrm{NO2}}$ = 2 at 293.15 K (− 0.239 ($n_{\mathrm{OH}}$ = 2); − 0.540 ($n_{\mathrm{OH}}$ = 3) [123]) with results for 1-alkanol + pentane systems at 298.15 K (− 0.264 ($n_{\mathrm{OH}}$ = 2); − 0.599 ($n_{\mathrm{OH}}$ = 3) [156]). Positive contributions to $\Delta\eta$, related to interactions between molecules, are also encountered in many others systems, as 1-alkanol + cyclohexylamine [157,158]. In fact, if interactions between unlike molecules are enough important, $\Delta\eta$ values may be positive. This is the case of 1-propanol, or 1-butanol + cyclohexylamine mixtures at 303.15 K (0.344 and 0.181 mPa·s, respectively [158]). The effect of the replacement of nitromethane by nitroethene in systems with a given 1-alkanol is not clear. It seems that $\Delta\eta$ values are slightly lower for methanol or ethanol + nitromethane mixtures (see above). This might be due to this nitroalkane is a better breaker of the alcohol self-association. More experimental work is needed in this field. Finally, we note that $\Delta\eta$ (nitroalkane) decreases when $n_{\mathrm{OH}}$ increases, which is the same behaviour observed for 1-alkanol + alkane, or + toluene systems. Therefore, it can be ascribed not only to a weakening of the interactions between unlike molecules caused by the OH group become more sterically hindered when $n_{\mathrm{OH}}$ increases, but also to the decreasing of the alcohol self-association at this condition.

*5.6 $S_{CC}(0)$ results.*

Inspection of Figures 9a and 9b allows state some interesting remarks. (i) We note that, for the studied systems, $S_{CC}(0)$ > 0.25. This means that homocoordination (i.e., interactions between like molecules) is the main feature for such mixtures. (ii) In the case of the nitrobenzene + heptane system, the $S_{CC}(0)$ curve shows a large maximum at 298.15 K (Figure 9a), a temperature close to the corresponding UCST (291.9 K [159]). As already mentioned, this is a typical behaviour shown by systems at temperatures in vicinity of the UCST, which are characterized by a strong homocoordination. (iii) The replacement of heptane by methanol leads to a large $S_{CC}(0)$ decrease (Figure 9a). Homocoordination becomes weaker due to the new methanol-nitrobenzene interactions created upon mixing. This newly demonstrates the existence of interactions between unlike molecules in 1-alkanol + nitroalkane mixtures. (iv) For nitromethane solutions, the maximum of the $S_{CC}(0)$ curves increases in the sequence: methanol > ethanol > 1-propanol > 1-butanol (Figure 9a). Interactions between like molecules become more relevant when $n_{\mathrm{OH}}$ increases, as then the system temperature is closer to the UCST. (v) For ethanol mixtures, $S_{CC}(0)$ changes in the order: nitromethane > nitroethane ≥ 1-nitropropane (Figure 9b). That is, interactions between like molecules are more relevant in nitromethane mixtures, and slightly more important in nitroethane systems than in those containing 1-nitropropane. (vi) Finally, the model consistently predicts the $S_{CC}(0)$ decrease of the 1-propanol + nitromethane system at 333.15 K, i.e., when the separation from the UCST increases. (Figures 9a and 9b).

### *5.7 Comparison between results from models and experimental data*

DISQUAC results on VLE, LLE, $H_{\mathrm{m}}^{\mathrm{E}}$, or $C_{\mathrm{p,m}}^{\mathrm{E}}$ (Tables 4-8) show that the model can be applied rather successfully over a wide range of temperature. Indeed, deviations between experimental and theoretical $H_{\mathrm{m}}^{\mathrm{E}}$ results for the systems methanol, ethanol, 1-propanol, or 1-butanol + nitromethane at 313.15 K are somewhat large. However, it must be underlined that the experimental data seem to be overestimated. Thus, the $C_{\mathrm{p,m}}^{\mathrm{E}}$ value, roughly evaluated from data at 298.15 K and 313.15 K, for the methanol + nitromethane mixture is 24.3 J·mol$^{-1}$·K$^{-1}$ [16], much higher than the value directly measured (9.3 J·mol$^{-1}$·K$^{-1}$ [17]). As a general trend, one can state that the larger differences between experimental results and DISQUAC calculations arise when the system temperature is close to the UCST. At this condition, the experimental $H_{\mathrm{m}}^{\mathrm{E}}$ and LLE curves become flattened, while the theoretical ones are more rounded. This is due to DISQUAC is a mean field theory and calculations are conducted assuming that the thermodynamic properties are analytical close to the critical temperature, when, really, they are expressed in terms of power laws. LLE curves are determined by means of DISQUAC assuming, erroneously, that $G_{\mathrm{m}}^{\mathrm{E}}$ is an analytical function close to the critical point. The instability of a system is given by $(\partial^2 G_{\mathrm{m}}^{\mathrm{M}} / \partial x_1^2)_{P,T}$ and represented by the critical exponent $\gamma > 1$ in the critical exponents theory [57]. In the framework of this theory, mean field models ($\gamma = 1$) provide LLE curves which are too high at the UCST and too low at the LCST [57] (lower critical solution temperature). For this reason, the $C_{\mathrm{sr,1}}^{\mathrm{DIS}}$ coefficients (s = a,h) must be ranged between certain limits in order to provide not very high calculated critical temperatures. This explains the difficulty in describing simultaneously VLE and LLE using the same interaction parameters and the slightly larger $\sigma_{\mathrm{r}}(P)$ values obtained in the case of 1-alkanol + nitromethane systems (Table 4). On the other hand, the critical exponent linked to the order parameter is, in mean field theories, $\beta$ = 0.5, and the derived LLE curves are more rounded close to the UCST, as fluctuations of the order parameter due to the sharp increase of the correlation length are not considered. For the nitrobenzene + *n*-alkane mixtures, there is an additional difficulty when describing the UCST variation with the number of C atoms of the alkane, as the experimental UCST values show a minimum for the heptane system (Table 6). That is, the solubility of mixtures involving pentane or hexane is lower. *N*-methylpyrrolidone systems show a similar trend, which has been explained by the phase rich in alkane is approaching to it gas-liquid critical point [160]. As already mentioned, $C_{\mathrm{p,m}}^{\mathrm{E}}$ curves of systems of the type polar compound + non-polar compound are W-shaped. In the case of 1-alkanol + nitromethane systems, small negative values are only encountered at low alcohol concentrations and at the temperatures closer to the UCST. This concentration dependence of $C_{\mathrm{p,m}}^{\mathrm{E}}$ is not properly described by the model (Figure 7). However, DISQUAC correctly predicts the decrease of $C_{\mathrm{p,m}}^{\mathrm{E}}$ when the difference (*T*-UCST) is increased (Table 8).

Comparison between DISQUAC and UNIFAC shows that DISQUAC describes more accurately the thermodynamic properties under consideration. Particularly, the temperature dependence of the mentioned properties is better represented by DISQUAC. For example, UNIFAC predicts UCST(nitrobenzene)/K = 280 (decane), 300 (hexadecane), which are poorer results than those provided by DISQUAC (Table 6). See also the results for $C_{\mathrm{p,m}}^{\mathrm{E}}$ (Table 8). DISQUAC also improves ERAS results on $H_{\mathrm{m}}^{\mathrm{E}}$ (Table 7) and $G_{\mathrm{m}}^{\mathrm{E}}$ (Figure S1, supplementary material). The most interesting result from the ERAS model is that deviations between experimental $H_{\mathrm{m}}^{\mathrm{E}}$ values and theoretical results increase with $n_{\mathrm{OH}}$ for nitromethane systems. That is, larger deviations are encountered when dipolar interactions become progressively more important. Particularly, we note that the theoretical $H_{\mathrm{m}}^{\mathrm{E}}$ curve of the 1-hexanol system is skewed towards low mole fractions of the alkanol (Figure 5), which suggests that effects related to the alcohol self-association are overestimated by the model. For the nitrobenzene solutions examined, the corresponding deviations are lower, as dipolar interactions are here comparatively less relevant. Nevertheless, we remark that the same type of asymmetry is encountered for the $H_{\mathrm{m}}^{\mathrm{E}}$ curve determined using ERAS for the 1-butanol mixture (Figure 6).

### *5.8 DISQUAC interaction parameters*

(i) The $C_{\mathrm{ar,1}}^{\mathrm{DIS}}$ coefficients in nitrobenzene + *n*-alkane mixtures change with the alkane size. As it has been explained above, this is needed to provide acceptable LLE results. The interaction parameters for nitrobenzene + hydrocarbon systems have been employed to predict $H_{\mathrm{m}}^{\mathrm{E}}$ values of related ternary mixtures. Calculations have been conducted using binary parameters only, i.e, neglecting ternary interactions. For the ternary mixtures nitrobenzene +

benzene + heptane ($N$ =17), or + cyclohexane ($N$ = 18) at 298.15 K [161], we have obtained, respectively, $dev(H_m^E)$ = 0.047 and 0.180. These so different results point out that the experimental values should be taken with caution, as we have demonstrated, along a number of works, that DISQUAC is a reliable tool to predict VLE and $H_m^E$ of ternary mixtures using only binary interaction parameters [162,163]. Regarding the mentioned systems, other theories do not provide better results. For example, the graph theory gives $dev(H_m^E)$ = 0.148 and 0.102 for the systems with heptane and cyclohexane, respectively [161]. The Flory model provides, in the same order, 0.205 and 0.035 for this magnitude [161], and from UNIFAC one obtains $dev(H_m^E)$ = 0.127 (heptane) and 0.316 (cyclohexane).

Finally, we have paid attention to the nitrobenzene + indane mixture, as solid-liquid equilibrium (SLE) data are available for this system [164]. Indane is important in the petrochemical industry and it is built by one aromatic ring and one aliphatic ring. Thus, there are three contacts in this solution: (b,c); (b,r) and (c,r). The experimental SLE data ($N$ = 23) are well represented by fitting only the $C_{cr,1}^{\mathrm{DIS}}$ coefficient (= – 1.55). Results (see Figure S3, supplementary material) were obtained using the needed physical constants of pure compounds available in the literature [164]. DISQUAC provides 0.005 and the Ideal Solubility Model 0.028 for $\sigma_r(T)$ (relative standard deviation for the temperature, defined similarly to $\sigma_r(P)$ (equation. 9)).

(ii) The $C_{\mathrm{hr},1}^{\mathrm{QUAC}}$ coefficients (l =1,2,3) are independent of the alcohol in systems with nitrobenzene. This behavior is also encountered in systems such as 1-alkanol + linear organic carbonate [165], or + *n*-alkanoate [166] or + *n*-alkanone [167] or + benzene, or + toluene [66].

(iii) The $C_{\mathrm{hr},1}^{\mathrm{QUAC}}$ coefficients (l = 1,2) for methanol, ethanol or 1-propanol + nitromethane systems are somewhat different from those of the remainder 1-alkanol + 1-nitroalkane mixtures. This merely reflects the difficulty in describing thermodynamics properties of systems including first members of homologous series [29]. Similar trends have been observed for 1-alkanol + *N,N*-dialkylamide [40], + *n*-alkanenitrile [41], + aniline [168], or + cyclic ether [69] mixtures.

### *5.9 ERAS interaction parameters*

The model provides rather reliable $H_m^E$ results (Table 7) using a consistent set of parameters (Table 3). Inspection of Table 3 allows conclude: (i) interactions between unlike molecules become less relevant when $n_{\mathrm{OH}}$ increases, as it is indicated by the corresponding $K_{\mathrm{AB}}$ decrease; (ii) the contribution to the excess functions arising from physical interactions is larger when $n_{\mathrm{OH}}$ increases. Table S2 (supplementary material) shows that the model gives correct $V_m^E$ ( $x_1$ = 0.5) values. However, it fails when describing $V_m^E$ ( $x_1$ ) (Figure S2, supplementary material). It means that the strong structural effects present in these solutions are not properly described by ERAS. In a previous study [149], better results on $V_m^E$ for 1-alkanol + nitromethane systems were obtained assuming that the nitroalkane is weakly self-associated, which is not justified. However, no result on $H_m^E$ was provided in that work [149].

## 6. Conclusions

From the existing database for the studied mixtures, it has been shown that, in alkane solutions, dipolar interactions between 1-nitroalkane molecules are weakened when $n_{NO2}$ increases. This is in agreement with the $\bar{\mu}(n_{NO2})$ variation. The aromaticity effect leads to interactions between nitrobenzene molecules are stronger than those involving the smaller 1-nitropropane. Dipolar interactions between like molecules are prevalent in systems with 1-alkanols, and become more important when $n_{OH}$ increases. At this condition, interactions between unlike molecules become weaker, as the OH group is more sterically hindered. Interactions between unlike molecules are stronger in systems with nitromethane than in nitrobenzene solutions. The replacement of nitromethane by nitroethane in systems with a given 1-alkanol leads to effects related with the alcohol self-association are more important. Permittivity data and the application of the Kirkwood-Fröhlich model for the $g_K$ determination show that the addition of 1-alkanol to a nitroalkane leads to cooperative effects, which increase the total effective dipole moment of the solution, in such way that the destruction of the existing structure in pure liquids is partially compensated. This effect is less important for larger $n_{OH}$ values.

## 7. List of symbols

| | |
|---|---|
| $C$ | interchange coefficient in DISQUAC |
| $C_p$ | heat capacity at constant pressure |
| $\Delta H$ | enthalpy of interaction |
| $g_K$ | Kirkwood's correlation factor (eq. 17) |
| $G$ | Gibbs energy |
| $H$ | enthalpy |
| $\Delta h_i^*$ | self-association enthalpy of component i |
| $\Delta h_{AB}^*$ | association enthalpy of component A with component B |
| $K_i$ | self-association constant of component i |
| $K_{AB}$ | association constant of component A with component B |
| $n_{NO2}$ | number of C atoms in 1-nitroalkane |
| $n_{OH}$ | number of C atoms in 1-alkanol |
| $P$ | pressure |
| $P_i^*$ | reduction parameter for pressure of the component i |
| $\bar{P}_i$ | reduced pressure of component i |
| $S$ | entropy |
| $S_{CC}(0)$ | concentration-concentration structure factor (eq. 7) |
| $T$ | temperature |
| $U_V$ | internal energy at constant volume |
| $V$ | volume |
| $V_{mi}^*$ | reduction parameter for molar volume of the component i |
| $\bar{V}_i$ | reduced volume of component i |
| $\Delta v_i^*$ | self-association volume of component i |
| $\Delta v_{AB}^*$ | association volume of component A with component B |
| $x$ | mole fraction in liquid phase |

*Greek letters*

| | |
|---|---|
| $\alpha_P$ | isobaric themal expansion coefficient |
| $\varepsilon_r$ | relative permittivity |
| $\eta$ | dynamic viscosity |
| $\kappa_T$ | isothermal compressibility |
| $\mu$ | dipole moment |
| $\bar{\mu}$ | effective dipole moment (eq. 11) |
| $\sigma_r$ | relative standard deviation (eq. 9) |
| $X_{12}$ | physical parameter in ERAS |

**Superscripts**

| | |
|---|---|
| E | excess property |

**Subscripts**

| | |
|---|---|
| i,j | compound in the mixture, (i, j =1,2) |
| m | molar property |
| s,t | type of contact surface in DISQUAC (s ≠ t = a ($CH_3$; $CH_2$); b ($C_6H_5$); c (c-$CH_2$); h, (OH); r (NO2) |

**Funding**

The authors gratefully acknowledge the financial support received from the Consejería de Educación y Cultura of Junta de Castilla y León, under Project BU034U16. F. Hevia gratefully acknowledges the grant received from the program 'Ayudas para la Formación de Profesorado Universitario (convocatoria 2014), de los subprogramas de Formación y de Movilidad incluidos en el Programa Estatal de Promoción del Talento y su Empleabilidad, en el marco del Plan Estatal de Investigación Científica y Técnica y de Innovación 2013-2016, de la Secretaría de Estado de Educación, Formación Profesional y Universidades, Ministerio de Educación, Cultura y Deporte, Gobierno de España'.

TABLE 1

Dispersive (DIS) and quasichemical (QUAC) interchange coefficients, $C_{\mathrm{sr},l}^{\mathrm{DIS}}$ and $C_{\mathrm{sr},l}^{\mathrm{QUAC}}$ ($l$ = 1, Gibbs energy; $l$ = 2, enthalpy; $l$ = 3, heat capacity) for (s,r) contacts[a] in nitrobenzene(1) + organic solvent(2) mixtures.

| Solvent | $C_{\mathrm{sr},1}^{\mathrm{DIS}}$ | $C_{\mathrm{sr},2}^{\mathrm{DIS}}$ | $C_{\mathrm{sr},3}^{\mathrm{DIS}}$ | $C_{\mathrm{sr},1}^{\mathrm{QUAC}}$ | $C_{\mathrm{sr},2}^{\mathrm{QUAC}}$ | $C_{\mathrm{sr},3}^{\mathrm{QUAC}}$ |
|---|---|---|---|---|---|---|
| | | contact: (b,r) | | | | |
| Benzene | 1.5 | 1.7 | − 1 | | | |
| | | contact: (b,r) | | | | |
| Toluene | − 2 | − 2.8 | − 1 | | | |
| | | contact: (a,r) | | | | |
| Pentane | − 0.32 | 0.12 | − 3 | 3.8 | 3.8 | 4 |
| Hexane | − 0.44 | 0.12 | − 3 | 3.8 | 3.8 | 4 |
| Heptane | − 0.535 | 0.12 | − 3 | 3.8 | 3.8 | 4 |
| Octane | − 0.60 | 0.12 | − 3 | 3.8 | 3.8 | 4 |
| Nonane | − 0.65 | 0.12 | − 2 | 3.8 | 3.8 | 4 |
| Decane | − 0.69 | 0.12 | − 1 | 3.8 | 3.8 | 4 |
| Dodecane | − 0.745 | 0.12 | − 1 | 3.8 | 3.8 | 4 |
| ≥ tridecane | − 0.783 | 0.12 | − 1 | 3.8 | 3.8 | 4 |
| | | contact: (c,r) | | | | |
| Cyclohexane | − 0.25[b] | 0.85 | − 3 | 3.8 | 3.8 | 4 |

[a]type a, aliphatic in *n*-alkanes or toluene; type, b, $C_6H_6$, or $C_6H_5$ in aromatic compounds considered (benzene, toluene or nitrobenzene); type c, c-$CH_2$ in cyclohexane; type r, $NO_2$ in nitrobenzene; [b]estimated value

TABLE 2

Dispersive (DIS) and quasichemical (QUAC) interchange coefficients, $C_{\mathrm{hr},l}^{\mathrm{DIS}}$ and $C_{\mathrm{hr},l}^{\mathrm{QUAC}}$ ($l$ = 1, Gibbs energy; $l$ = 2, enthalpy; $l$ = 3, heat capacity) for (h,r) contacts[a] in 1-alkanol(1) + 1-nitroalkane(2) or + nitrobenzene(2) mixtures.

| $n_{\mathrm{OH}}$ [b] | $C_{\mathrm{hr},1}^{\mathrm{DIS}}$ | $C_{\mathrm{hr},2}^{\mathrm{DIS}}$ | $C_{\mathrm{hr},3}^{\mathrm{DIS}}$ | $C_{\mathrm{hr},1}^{\mathrm{QUAC}}$ | $C_{\mathrm{hr},2}^{\mathrm{QUAC}}$ | $C_{\mathrm{hr},3}^{\mathrm{QUAC}}$ |
|---|---|---|---|---|---|---|
| | | | 1-alkanol + nitromethane | | | |
| 1 | 11 | 10.8 | 2 | − 2.5 | − 1.8 | 5 |
| 2 | 7.8 | 10.8 | 16 | − 0.35 | − 1.6 | 5 |
| 3 | 7.8 | 10.8 | 16 | − 0.1 | − 1 | 5 |
| 4 | 0.6 | 6 | 23 | 5.2 | − 1 | 5 |
| 6 | 2 | 6 | 40 | 5.2 | − 1 | 5 |
| 8 | 4.5[c] | 6 | 40 | 5.2 | − 1 | 5 |
| 10 | 7.8 | 6 | 40 | 5.2 | − 1 | 5 |
| 12 | 13 | 6 | 40 | 5.2 | − 1 | 5 |
| 15 | 21.6 | 6 | 40 | 5.2 | − 1 | 5 |
| | | | 1-alkanol + ≥ nitroethane | | | |
| 1 | − 0.3 | 6 | 2 | 5.2 | − 1 | 5 |
| 2 | − 0.3 | 6 | 16 | 5.2 | − 1 | 5 |
| 3 | 0.2 | 6 | 16 | 5.2 | − 1 | 5 |
| 4 | 0.85[c] | 6 | 16 | 5.2 | − 1 | 5 |
| 6 | 2.6[c] | 6 | 40 | 5.2 | − 1 | 5 |
| 8 | 5[c] | 6 | 40 | 5.2 | − 1 | 5 |
| 10 | 8.8 | 6 | 40 | 5.2 | − 1 | 5 |
| 12 | 14[c] | 6 | 40 | 5.2 | − 1 | 5 |
| 15 | 23[c] | 6 | 40 | 5.2 | − 1 | 5 |
| | | | 1-alkanol + nitrobenzene | | | |
| 1 | 0.12 | 2 | | 7 | 6.8 | |
| 2 | 1[c] | 2.3 | | 7 | 6.8 | |
| 3 | 2[c] | 3.5 | | 7 | 6.8 | |
| 4 | 3[c] | 3.5 | | 7 | 6.8 | |
| 7 | 6 | 3.5 | | 7 | 6.8 | |

[a]type h, OH in 1-alkanols, type r, $NO_2$ 1-nitroalkanes or in nitrobenzene; [b] $n_{\mathrm{OH}}$ is the number of C atoms in the 1-alkanol.

TABLE 3

ERAS parameters[a] for 1-alkanol(1) + 1-nitroalkane(2) or + nitrobenzene(2) mixtures at 298.15 K.

| System | $K_{AB}$ | $\Delta h^*_{AB}$ /kJ·mol$^{-1}$ | $\Delta v^*_{AB}$ / cm$^3$· mol$^{-1}$ | $X_{AB}$ /J· cm$^3$ |
|---|---|---|---|---|
| Methanol + nitromethane | 118 | – 15 | – 7.5 | 16 |
| ethanol + nitromethane | 82 | – 15 | – 7.5 | 32 |
| 1-propanol + nitromethane | 82 | – 15 | – 7.5 | 45 |
| 1-butanol + nitromethane | 45 | – 15 | – 9.5 | 68 |
| 1-hexanol + nitromethane[b] | 25 | – 15 | – 9.5 | 110 |
| Methanol + nitrobenzene | 80 | – 15 | – 7. | 16 |
| ethanol + nitrobenzene | 60 | – 15 | – 6.1 | 22 |
| 1-propanol + nitrobenzene | 60 | – 15 | – 6.1 | 36 |
| 1-butanol + nitrobenzene | 60 | – 15 | – 6.6 | 36 |

[a] $K_{AB}$, association constant of component A with component B; $\Delta h^*_{AB}$, association enthalpy of component A with component B; $\Delta v^*_{AB}$, association volume of component A with component B; $X_{AB}$, physical parameter; [b]system at 313.15 K

TABLE 4

Molar excess Gibbs energies, $G_m^E$, at equimolar composition and temperature $T$ for nitrobenzene(1) + organic solvent(2) mixtures or for 1-alkanol(1) + 1-nitroalkane(2) systems.

| System | $T$/K | $N$[a] | $G_m^E$ /J mol$^{-1}$ | | $\sigma_r(P)$ [b] | | | Ref. |
|---|---|---|---|---|---|---|---|---|
| | | | exp[c] | DQ[d] | exp[c] | DQ[d] | UNIF[e] | |
| Nitrobenzene + $C_6H_{12}$ | 353.15 | 7 | 1160 | 1174 | 0.004 | 0.019 | 0.087 | 169 |
| Nitrobenzene + $C_7H_8$ | 373.15 | 6 | 334 | 335 | 0.003 | 0.015 | 0.024 | 170 |
| Methanol + nitromethane | 298.15 | 13 | 1040 | 1036 | 0.002 | 0.034 | 0.030 | 14 |
| | | 9 | 1030 | | 0.003 | 0.025 | 0.024 | 171 |
| | 348.15 | 13 | 982 | 968 | 0.001 | 0.016 | 0.024 | 14 |
| | 388.24 | 13 | 971 | 893 | 0.007 | 0.025 | 0.032 | 14 |
| ethanol + nitromethane | 298.15 | 13 | 1170 | 1179 | 0.002 | 0.013 | 0.010 | 14 |
| | 348.15 | 13 | 1040 | 1029 | 0.0005 | 0.012 | 0.008 | 14 |
| | 398.17 | 13 | 911 | 812 | 0.0003 | 0.035 | 0.007 | 14 |
| 1-propanol + nitromethane | 333.15 | 23 | 1170 | 1021 | 0.001 | 0.055 | 0.017 | 15 |
| Methanol + nitroethane | 298.15 | 9 | 1000 | 1009 | 0.002 | 0.026 | 0.041 | 171 |
| | 342.50 | | 974[f] | 947 | | | | 172 |
| | | | | (932)[e] | | | | |
| ethanol + nitroethane | 354.23 | | 857[g] | 863 | | | | 172 |
| | | | | (864)[e] | | | | |
| 1-propanol + nitroethane | 368.72 | | 800[h] | 814 | | | | 172 |
| | | | | (808)[e] | | | | |
| Methanol + nitrobenzene | 323.35 | 16 | 1270 | 1251 | 0.005 | 0.035 | | 173 |
| | 423.15 | 16 | 992 | 998 | 0.015 | 0.031 | | 173 |
| 1-heptanol + nitrobenzene | 450.65 | | 1329[i] | 1329 | | | | 174 |
| | | | | (778)[e] | | | | |

[a]number of experimental data; [b]relative standard deviation for pressure (equation 9); [c]experimental result; [d] DISQUAC result using interaction parameters listed in Tables 1 and 2; [e]UNIFAC result using interaction parameters from the literature [32]; [f] $x_1 = 0.454$; [g] $x_1 = 0.486$; [h] $x_1 = 0.504$; [i] $x_1 = 0.521$

TABLE 5

Coordinates[a] of the azeotropes for 1-alkanol(1) + 1-nitroalkane(2), or + nitrobenzene(2) mixtures.

| System | $T_{az}$ /K | | $P_{az}$ /kPa | | $x_{1az}$ | | Ref. |
|---|---|---|---|---|---|---|---|
| | Exp[b] | DQ[c] | Exp[b] | DQ[c] | Exp[b] | DQ[c] | |
| Methanol + nitromethane | 348.15 | 348.15 | 151.91 | 152.0 | 0.952 | 0.954 | 14 |
| Ethanol + nitromethane | 348.15 | 348.15 | 97.57 | 95.92 | 0.761 | 0.768 | 14 |
| 1-propanol + nitromethane | 333.15 | 333.15 | 33.68 | 32.20 | 0.448 | 0.406 | 15 |
| Ethanol + nitroethane | 351.21 | 351.21 | 101.32 | 100.52 | 0.931 | 0.970 | 172 |
| 1-propanol + nitroethane | 367.88 | 367.88 | 101.32 | 99.25 | 0.743 | 0.761 | 172 |
| 1-heptanol + nitrobenzene | 448.15 | 448.15 | 96.5 | 100.45 | 0.844 | 0.837 | 174 |

[a] $T_{az}$ , temperature; $P_{az}$ , pressure; $x_{1az}$ , mole fraction; [b]experimental result, [c]DISQUAC result using interaction parameters listed in Tables 1 and 2

TABLE 6

Coordinates of the critical points[a] for nitrobenzene(1) + *n*-alkane(2) mixtures and for 1-alkanol(1) + 1-nitroalkane(2) systems

| | $x_{1c}$ | | $T_c$ /K | | Ref |
|---|---|---|---|---|---|
| | Exp[b] | DQ[c] | Exp[b] | DQ[c] | |
| | | Nitrobenzene + *n*-alkane | | | |
| *n*-$C_5$ | 0.386 | 0.345 | 297.1 | 291.2 | 76 |
| *n*-$C_6$ | 0.428 | 0.392 | 293.1 | 292.7 | 159 |
| *n*-$C_7$ | 0.471 | 0.439 | 291.9 | 293.3 | 159 |
| *n*-$C_8$ | 0.505 | 0.481 | 293.1 | 294.9 | 75 |
| *n*-$C_9$ | 0.544 | 0.521 | 294.2 | 296.5 | 175 |
| *n*-$C_{10}$ | 0.574 | 0.559 | 296.0 | 298.2 | 76 |
| *n*-$C_{12}$ | 0.630 | 0.621 | 300.4 | 301.6 | 176 |
| *n*-$C_{14}$ | 0.676 | 0.681 | 304.9 | 305.0 | 177 |
| *n*-$C_{16}$ | 0.713 | 0.736 | 309.7 | 312.4 | 23 |
| | | 1-alkanol + nitromethane | | | |
| 1-butanol | 0.405 | 0.393 | 291.14 | 293.9 | 10,11 |
| | 0.418 | | 290.35 | | 19 |
| 1-hexanol | 0.352 | 0.292 | 308.65 | 309.7 | 10,11 |
| | 0.320 | | 308.75 | | 100 |
| 1-dodecanol | 0.179 | 0.145 | 341.32 | 343.2 | 10,11 |
| 1-pentadecanol | 0.136 | 0.111 | 352.65 | 354.7 | 11,12 |
| | | 1-alkanol + nitroethane | | | |
| 1-decanol | 0.240 | 0.234 | 294.1 | 296.8 | 11,13 |

[a]critical composition, $x_{1c}$ ; UCST, $T_c$ ; [b]experimental value; [c]DISQUAC results using interaction parameters listed in Tables 1 and 2

TABLE 7

Molar excess enthalpies, $H_m^E$, at equimolar composition and temperature $T$ for nitrobenzene(1) + organic solvent(2) mixtures or for 1-alkanol(1) + 1-nitroalkane(2), or + nitrobenzene(2) systems.

| System | $T$/K | $N$[a] | $H_m^E$/J mol$^{-1}$ | | $dev(H_m^E)$ [b] | | | Ref. |
|---|---|---|---|---|---|---|---|---|
| | | | exp[c] | DQ[d] | exp[c] | DQ[d] | UNIF/ ERAS[e] | |
| Nitrobenzene + benzene | 293.15 | 17 | 261 | 262 | 0.006 | 0.020 | 0.142[f] | 60 |
| Nitrobenzene + toluene | 293.15 | 17 | 221 | 217 | 0.012 | 0.030 | 0.118[f] | 60 |
| Nitrobenzene + hexane | 298.15 | 12 | 1447 | 1467 | 0.004 | 0.020 | 0.209[f] | 78 |
| | 323.15 | 12 | 1498 | 1500 | 0.002 | 0.009 | 0.084[f] | 78 |
| Nitrobenzene + $C_6H_{12}$ | 293.15 | 9 | 1654 | 1667 | 0.027 | 0.024 | 0.326[f] | 77 |
| Methanol + nitromethane | 298.15 | 19 | 1265 | 1262 | 0.003 | 0.068 | 0.035[f]/0.058[g] | 16 |
| | | 6 | 1268 | | 0.017 | 0.087 | 0.041[f] | 128 |
| | 313.15 | 19 | 1629 | 1387 | 0.002 | 0.149 | 0.132[f] | 16 |
| ethanol + nitromethane | 298.15 | 19 | 1632 | 1715 | 0.002 | 0.045 | 0.362[f]/0.074[g] | 16 |
| | 313.15 | 19 | 2086 | 1959 | 0.003 | 0.104 | 0.104[f] | 16 |
| 1-propanol + nitromethane | 298.15 | 19 | 1911 | 2027 | 0.002 | 0.042 | 0.323[f]/0.103[g] | 16 |
| | 313.15 | 19 | 2644 | 2256 | 0.004 | 0.132 | 0.059[f] | 16 |
| 1-butanol + nitromethane | 294.15 | 14 | 1937 | 2126 | 0.007 | 0.088 | 0.402[f] | 16 |
| | 295.15 | 18 | 2095 | 2147 | 0.009 | 0.095 | 0.326[f] | 178 |
| | 298.15 | 19 | 2131 | 2211 | 0.003 | 0.052 | 0.313[f]/0.112[g] | 16 |
| | | 18 | 2220 | | 0.005 | 0.095 | 0.261[f] | 178 |
| | 303.15 | 18 | 2312 | 2315 | 0.005 | 0.086 | 0.209[f] | 178 |
| | 313.15 | 19 | 2810 | 2518 | 0.002 | 0.088 | 0.011[f] | 16 |
| 1-hexanol + nitromethane | 313.15 | 15 | 2781 | 2848 | 0.002 | 0.027 | 0.120[f]/0.127[g] | 114 |
| Methanol + nitrobenzene | 298.15 | 8 | 1109 | 1158 | 0.024 | 0.126 | 0.056[g] | 128 |
| ethanol + nitrobenzene | 298.15 | 9 | 1430 | 1446 | 0.011 | 0.091 | 0.365[f]/0.062[g] | 129 |
| 1-propanol + nitrobenzene | 298.15 | 10 | 1807 | 1839 | 0.007 | 0.049 | 0.402[f]/0.076[g] | 129 |
| 1-butanol + nitrobenzene | 298.15 | 9 | 1946 | 1977 | 0.006 | 0.056 | 0.375[f]/0.077[g] | 129 |

[a]number of experimental data; [b]equation (10); [c]experimental result; [d]DISQUAC result using interaction parameters listed in Tables 1 and 2; [e]UNIFAC or ERAS result; [f]UNIFAC result obtained with interaction parameters from literature [32]; [g]ERAS result using parameters listed in Table 3

TABLE 8

Isobaric molar excess heat capacities, $C_{pm}^{E}$ , at equimolar composition and temperature $T$ for 1-alkanol(1) + 1-nitroalkane(2) systems.

| System | $T$/K | $C_{pm}^{E}$ /J·mol$^{-1}$·K$^{-1}$ | | | Ref |
|---|---|---|---|---|---|
| | | Exp[a] | DQ[b] | UNIF[c] | |
| Methanol + nitromethane | 298.15 | 9.3 | 8.8 | 1.9 | 17 |
| 1-propanol + nitromethane | 288.15 | 19.1 | 17.8 | − 2.9 | 18 |
| | 298.15 | 16.6 | 16.4 | − 7.5 | 18 |
| | 308.15 | 14.7 | 14.9 | − 10.9 | 18 |
| 1-butanol + nitromethane | 293.15 | 26.0 | 21.4 | − 8.2 | 19 |
| | 298.15 | 20.9 | 21.0 | − 10.6 | 19 |
| | 308.15 | 16.3 | 20.3 | − 14.6 | 19 |
| Ethanol + 1-nitropropane | 298.15 | 14.9 | 14.7 | 1.4 | 20 |

[a]experimental result; [b]DISQUAC result using interaction parameters listed in Tables 1 and 2
[c]UNIFAC result obtained with interaction parameters from literature [32]

TABLE 9

Physical constants of pure nitroalkanes[a]

| Compound | $P_c$ /bar | $T_c$ /K | $V_m$ /cm$^3$·mol$^{-1}$ | $\mu$ /D | $\mu_{eff}$ |
|---|---|---|---|---|---|
| Nitromethane | 63.1 | 588 | 53.96 | 3.56 | 1.855 |
| Nitroethane | 51.2[b] | 557 | 71.86 | 3.60 | 1.625 |
| 1-nitropropane | 44.8[b] | 675.2 | 89.44 | 3.59 | 1.453 |
| Nitrobenzene | 46.5[b] | 732 | 102.74 | 4.0 | 1.510 |

[a] $P_c$ , critical pressure; $T_c$ , critical temperature; $V_m$ , molar volume; $\mu$ , dipole moment; $\mu_{eff}$ , effective dipole moment (equation 11), data taken from reference [1], except when indicated;
[b]from the application of Joback's method [180]

TABLE 10

Partial molar excess enthalpies,[a] $H_1^{E,\infty}$, at 298.15 K at 0.1 MPa for nitromethane(1) or nitrobenzene(1) + alkane(2) or for 1-alkanol(1) + nitroalkane(2) mixtures, and hydrogen bond enthalpies, $\Delta H_{OH\text{-}NO2}$, for 1-alkanol(1) + nitromethane(2), or + nitrobenzene(2) systems.

| System | $H_1^{E,\infty}$ /kJ·mol$^{-1}$ | $\Delta H_{OH\text{-}NO2}$ /kJ·mol$^{-1}$ |
|---|---|---|
| Nitromethane(1) + cyclohexane(2) | 16.5 [179] | |
| Nitrobenzene(1) + hexane(2) | 9.53 [78] | |
| Methanol(1) + nitromethane (2) | 9.29 [179] | −30.4 |
| Ethanol(1) + nitromethane(2) | 11.97 [179] | −27.7 |
| 1-propanol(1) + nitromethane(2) | 13.68 [179] | −26.0 |
| 1-butanol(1) + nitromethane(2) | 16.00 [179] | −23.7 |
| Methanol(1) + nitrobenzene (2) | 9.19 [128] | −23.5 |
| Ethanol(1) + nitrobenzene(2) | 11.75 [129] | −21.0 |
| 1-propanol(1) + nitrobenzene(2) | 13.63 [129] | −19.1 |
| 1-butanol(1) + nitrobenzene(2) | 14.72 [129] | −18.0 |

[a]values obtained from $H_m^E$ data over the whole concentration range

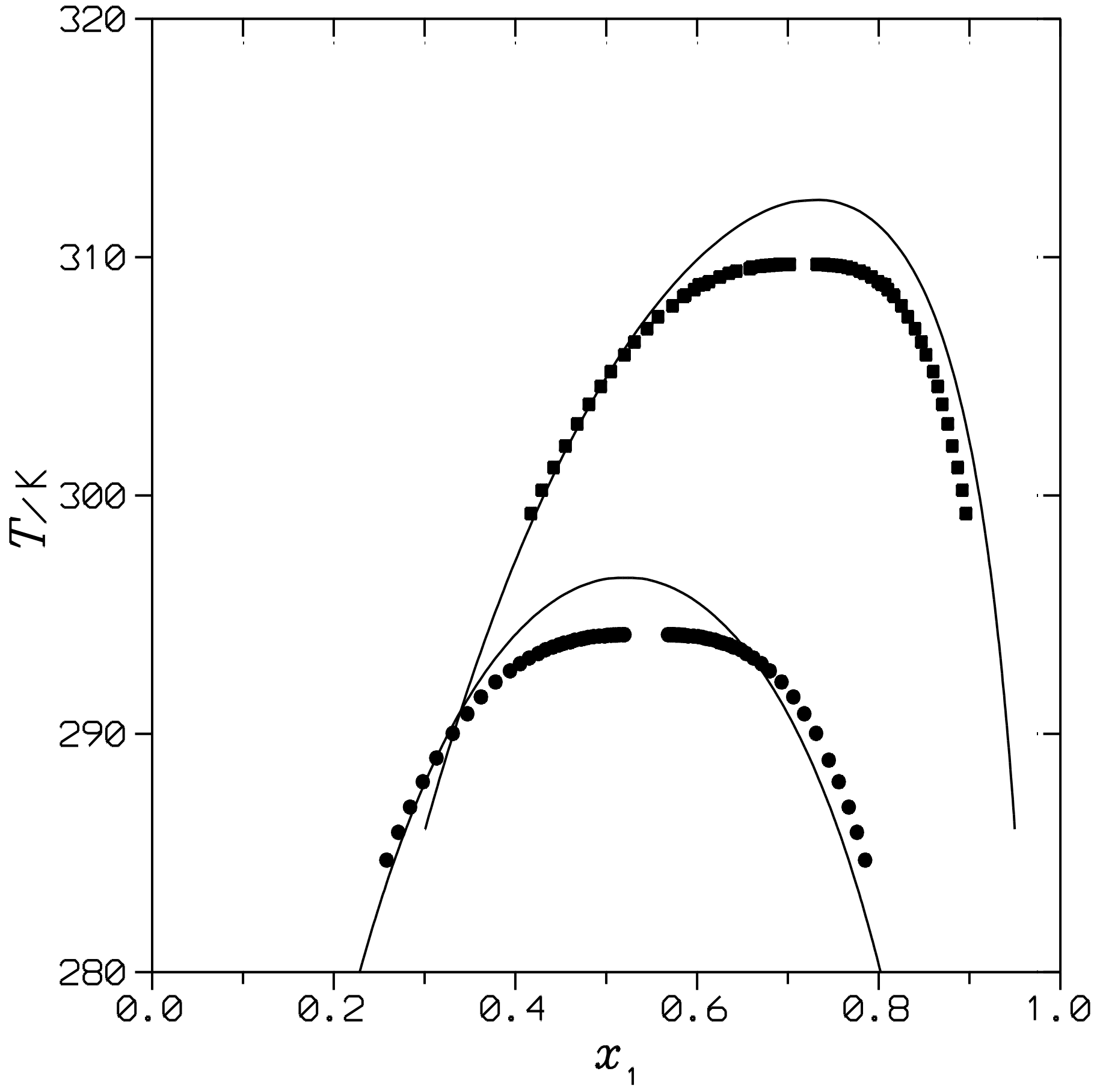

Figure 1 LLE for nitrobenzene(1) + nonane(2) (●, [175]), or + hexadecane (■, [23]) mixtures. Solid lines, DISQUAC calculations.

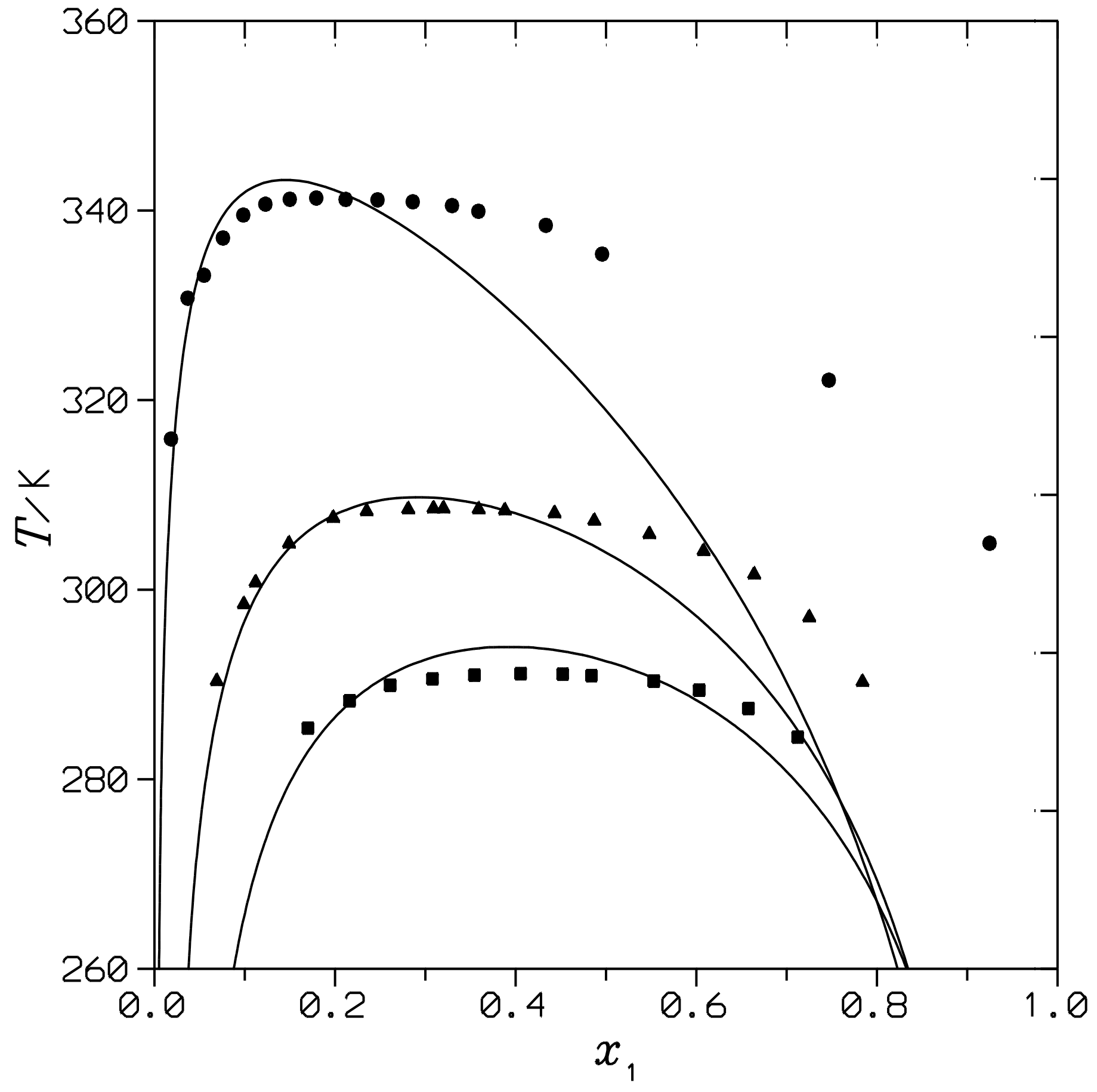

Figure 2 LLE for 1-alkanol(1) + nitromethane(2) mixtures. Points, experimental results: (■), 1-butanol [10,11]; (▲), 1-hexanol [100]; (●), 1-dodecanol [10,11]. Solid lines, DISQUAC calculations.

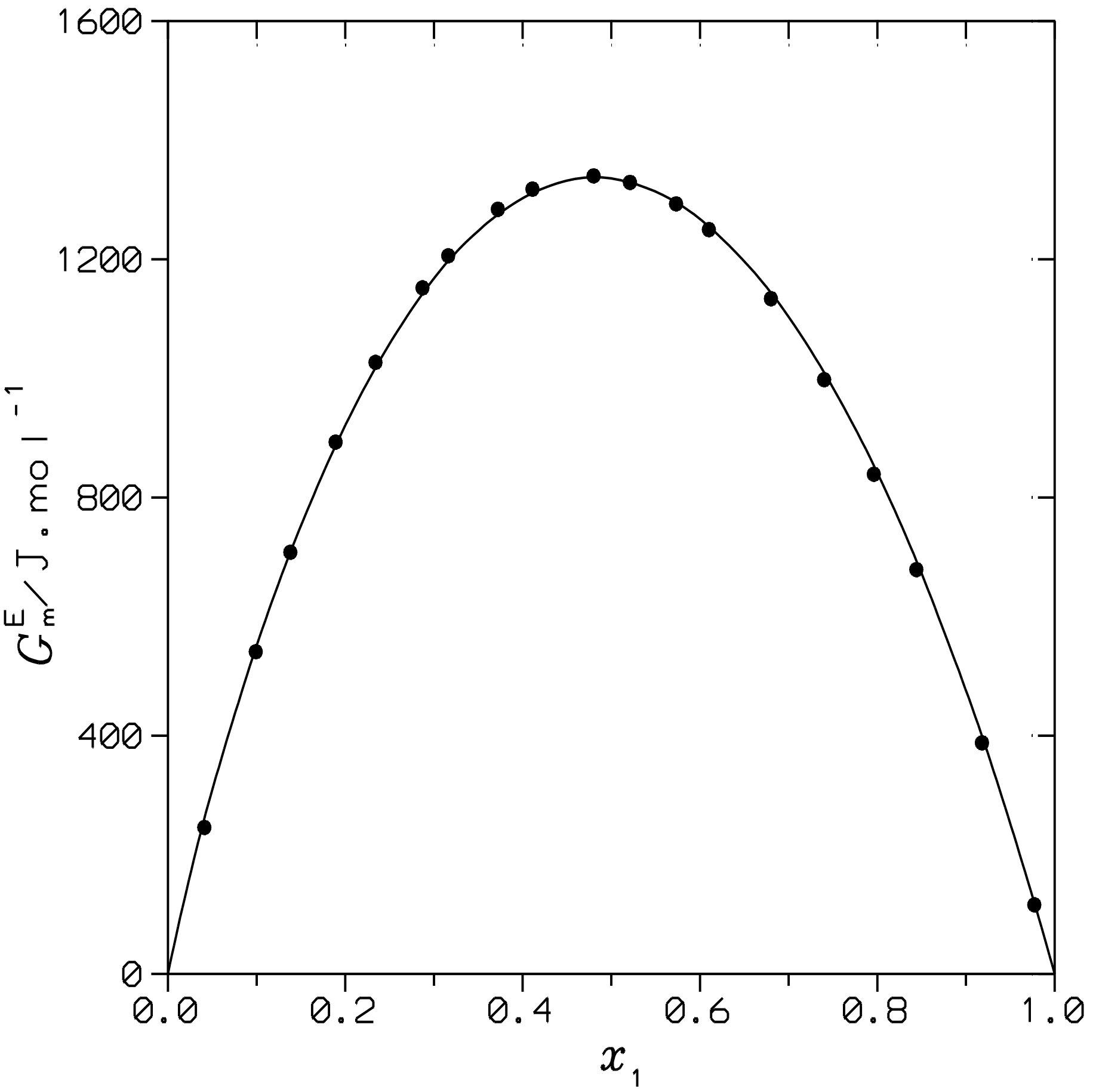

Figure 3 $G_m^E$ for the 1-heptanol(1) + nitrobenzene(2) mixture over the temperature range (448.15-470.15) K. Points, experimental results [174]. Solid line, DISQUAC calculations.

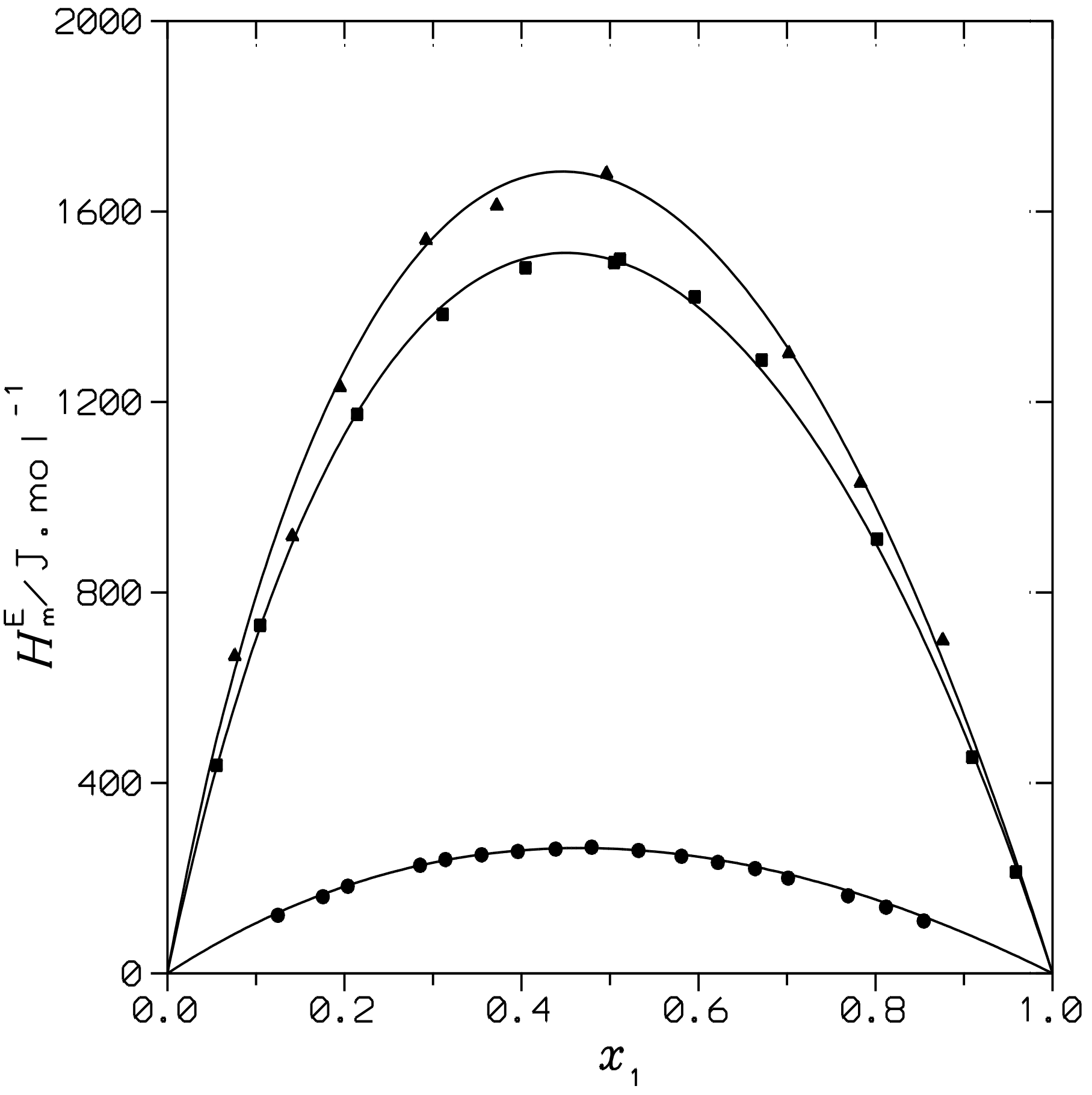

Figure 4 $H_m^E$ for nitrobenzene(1) + organic solvent(2) mixtures. Points, experimental results: (●), benzene ($T$ = 293.15 K, [60]); (■), hexane ($T$ = 323.15 K, [78]); (▲), cyclohexane $T$ = 293.15 K, [77]). Solid lines, DISQUAC calculations.

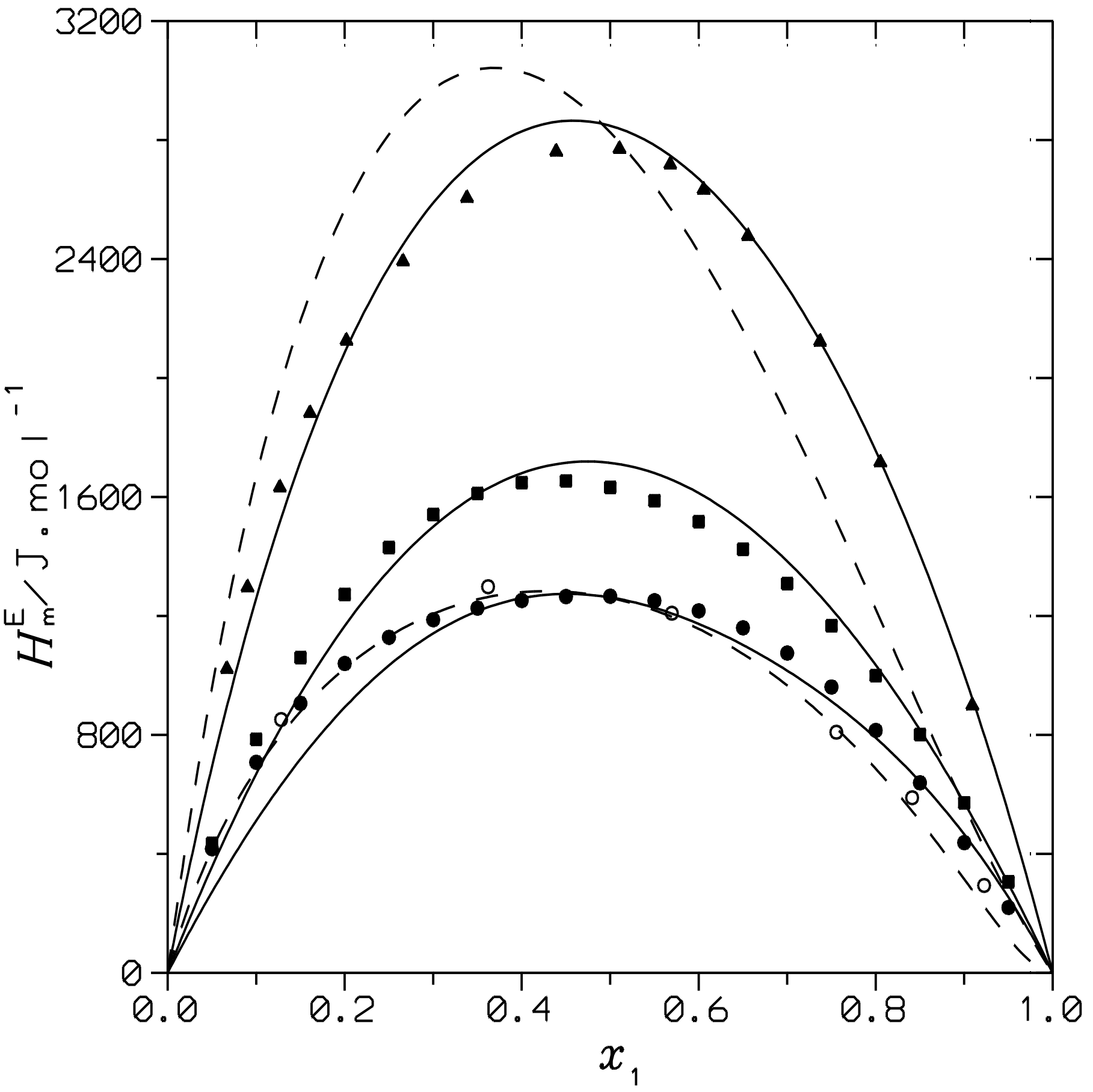

Figure 5 $H_m^E$ for 1-alkanol(1) + nitromethane(2) mixtures. Points, experimental values: (●) [16]; (O) [128], methanol ($T$ = 298.15 K); (■), ethanol ($T$ = 298.15 K, [16]); (▲), 1-hexanol ($T$ = 313.15 K, [114]). Solid lines, DISQUAC calculations. Dashed lines, ERAS results.

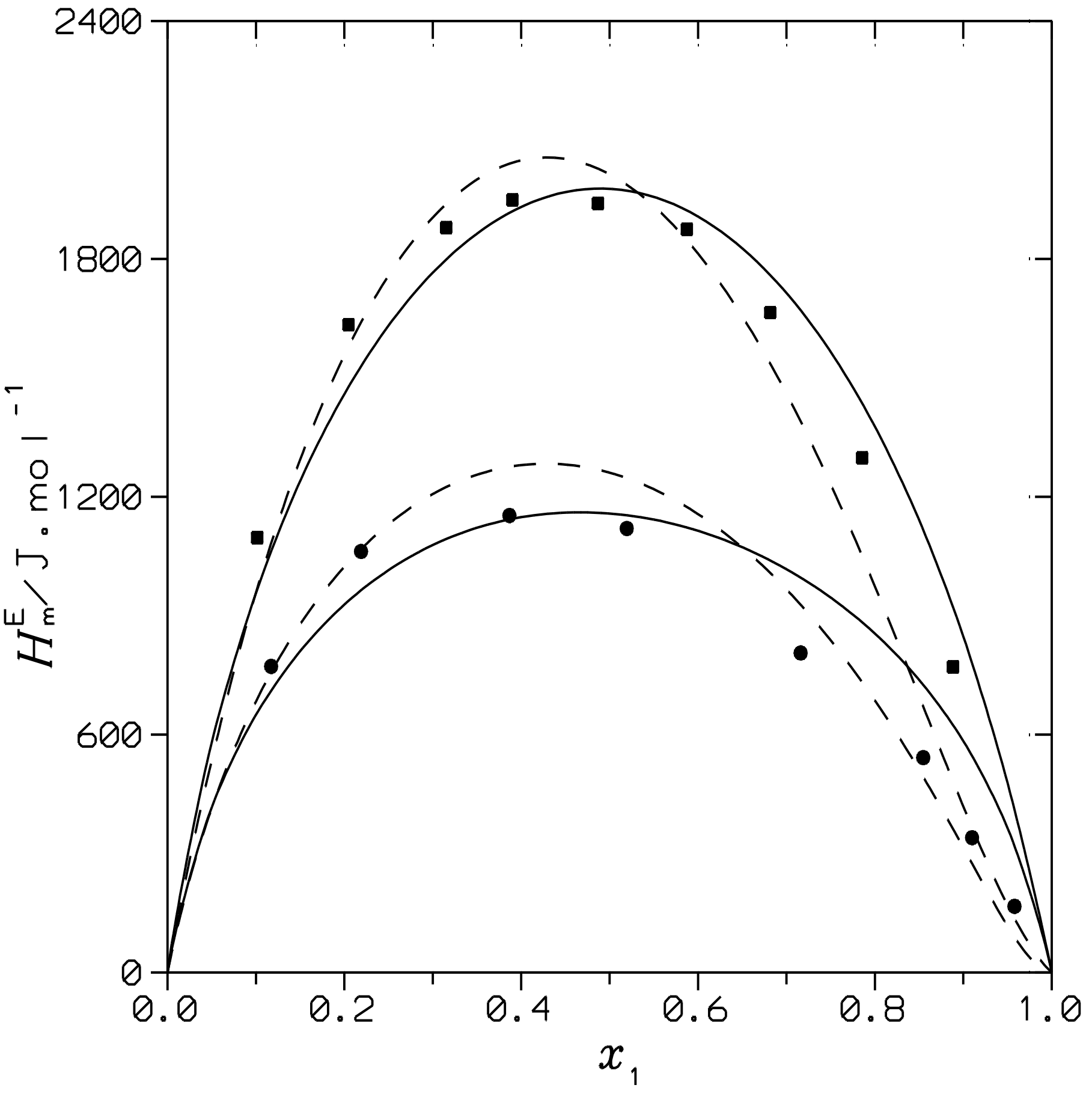

Figure 6 $H_m^E$ for 1-alkanol(1) + nitrobenzene(2) mixtures at 298.15 K. Points, experimental values [129]: (●), methanol; (■), 1-butanol. Solid lines, DISQUAC calculations. Dashed lines, ERAS results.

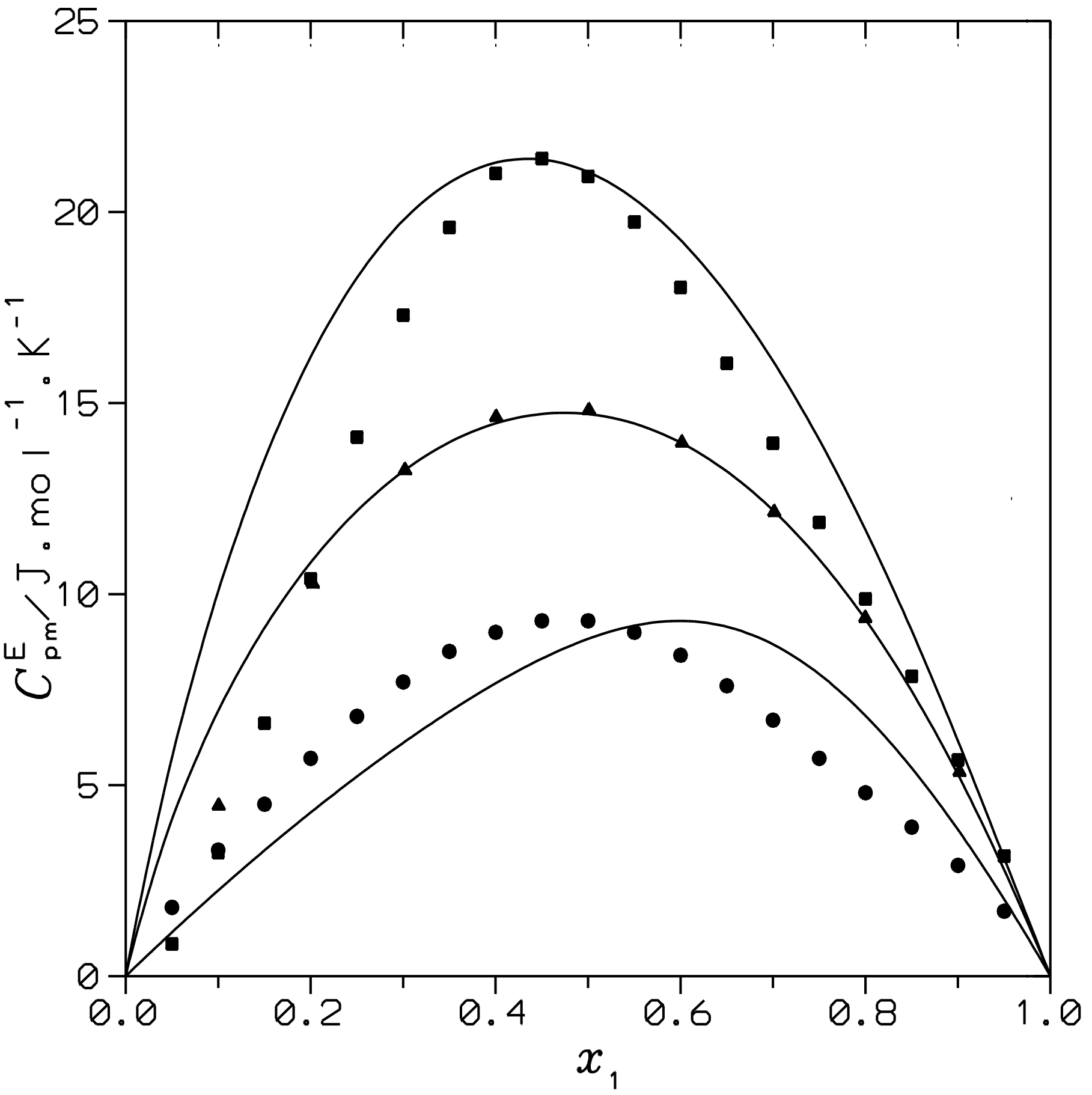

Figure 7 $C^{E}_{pm}$ for 1-alkanol(1) + 1-nitroalkane(2) mixtures at 298.15 K. Points, experimental values: (●), methanol + nitromethane [17]; (■),1-butanol + nitromethane [19]; (▲), ethanol + 1-nitropropane [20]. Solid lines, DISQUAC calculations.

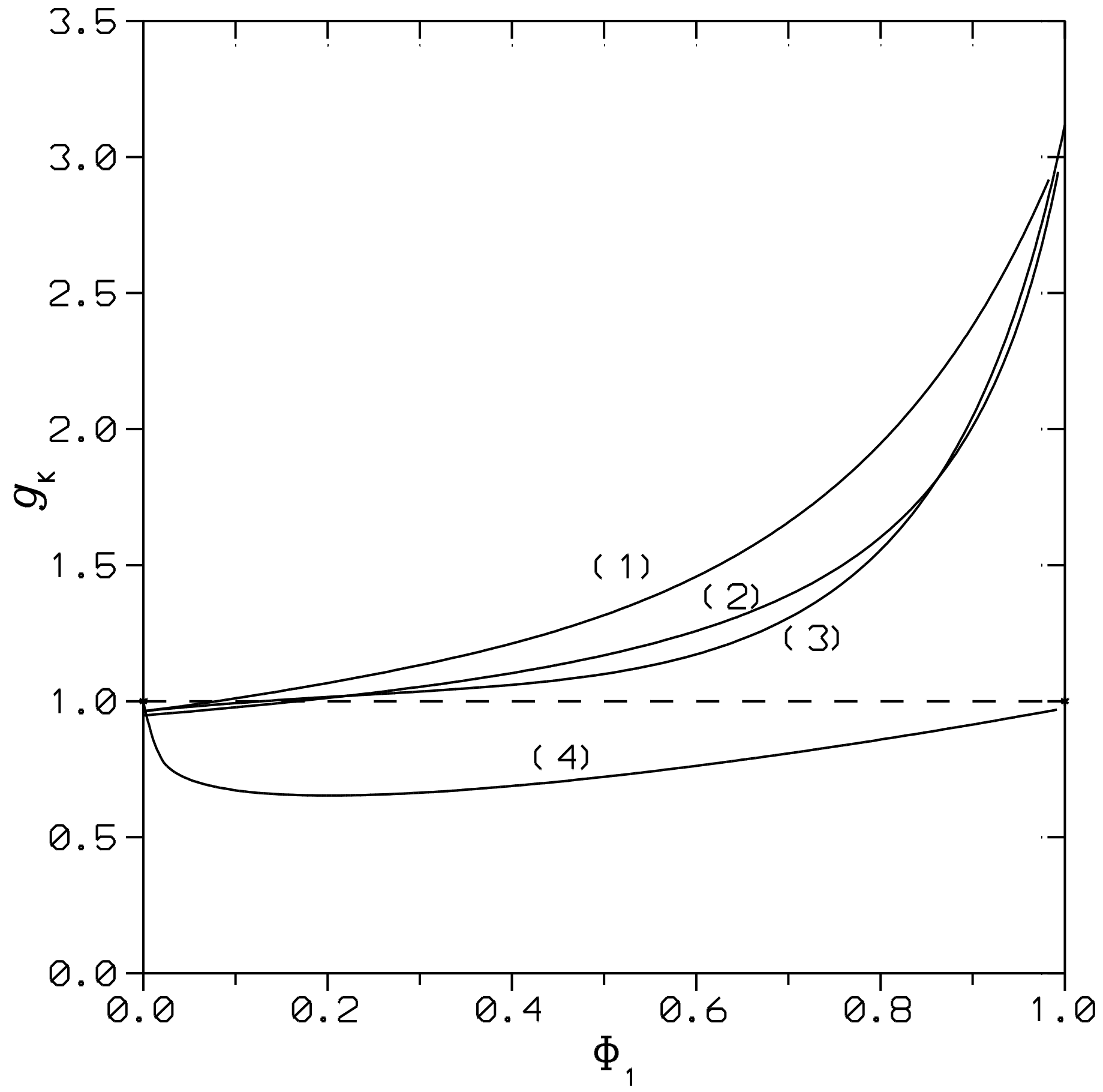

Figure 8 Kirkwood's correlation factor, $g_K$, vs. volume fraction for the mixtures: (1), ethanol(1) + nitrobenzene(2); (2), 1-butanol(1) + nitrobenzene(2) ($T$ = 298.15 K); (3), 1-heptanol(1) + nitrobenzene(2) ($T$ = 293.15 K); (4) nitrobenzene(1) + benzene(2) ($T$ = 298.15 K). For references needed for calculations see text.

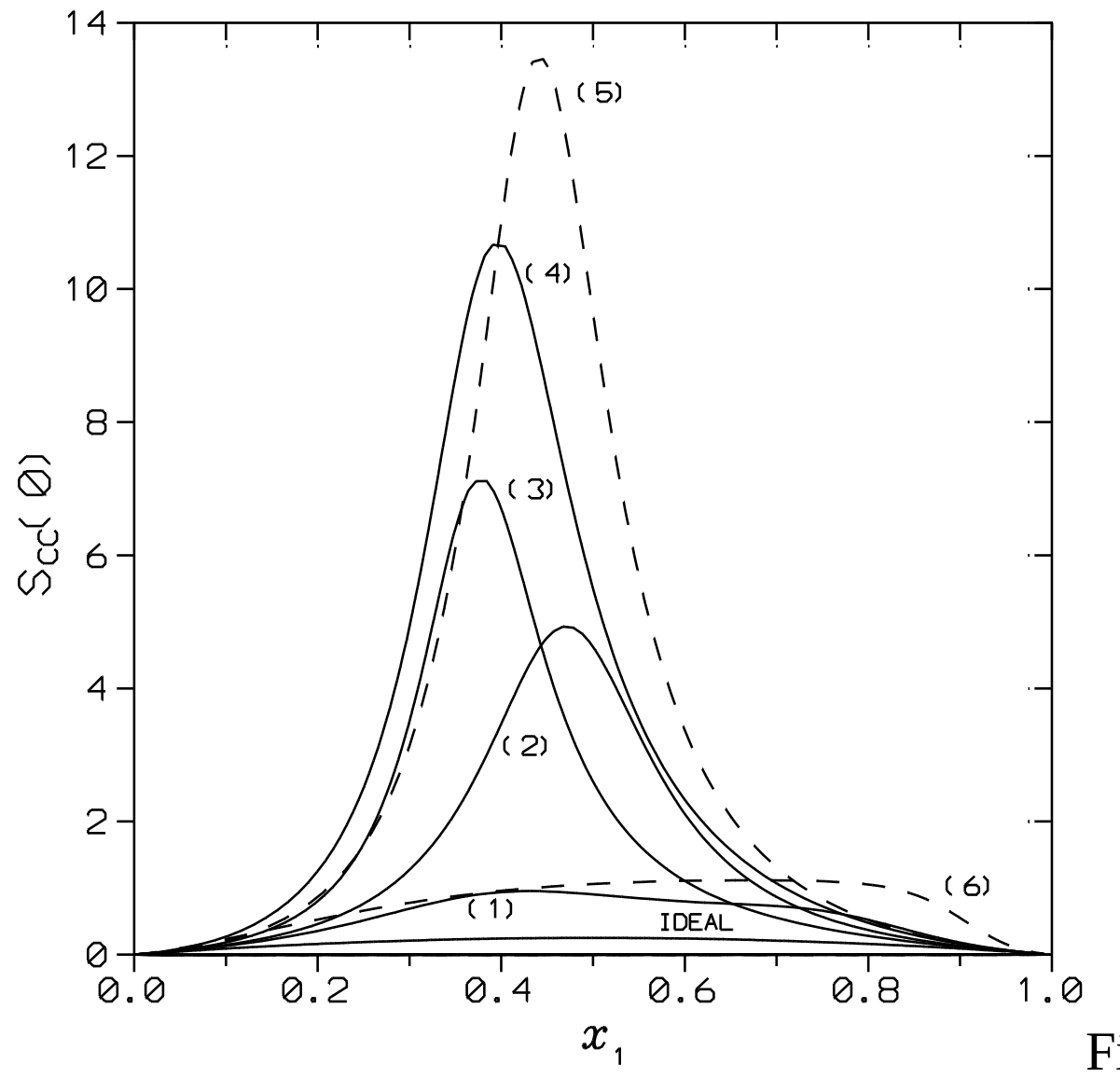

Figure 9a DISQUAC calculations on $S_{CC}(0)$ at 298.15 K for the mixtures: (1), methanol(1) + nitromethane(2); (2), ethanol(1) + nitromethane(2); (3), 1-propanol(1) + nitromethane(2); (4), 1-butanol(1) + nitromethane(2); (5), nitrobenzene(1) + heptane(2); (6), methanol(1) + nitrobenzene(2)

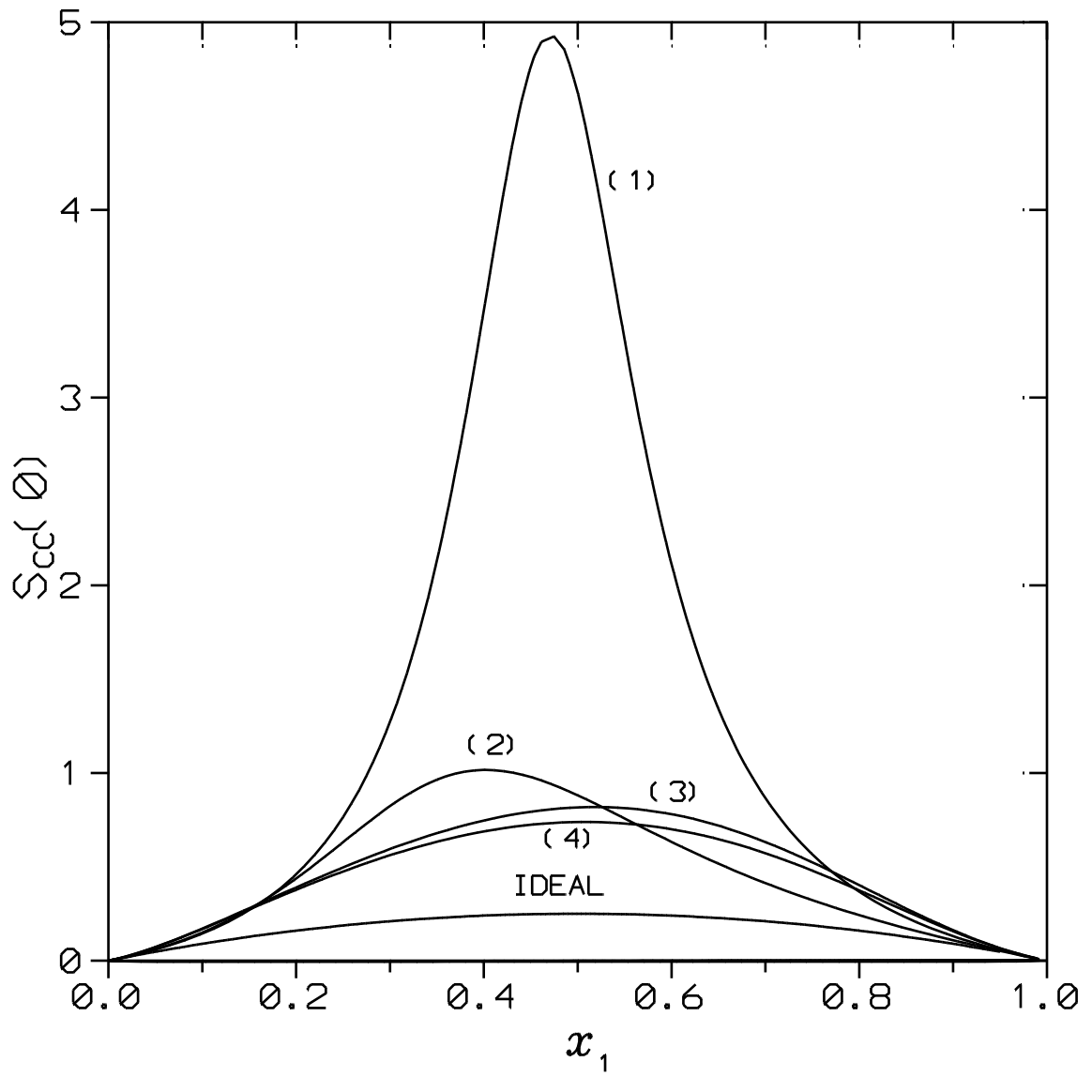

Figure 9b DISQUAC calculations on $S_{CC}(0)$ for the mixtures: (1), ethanol(1) + nitromethane(2); (3), ethanol(1) + nitroethane(2); (4), ethanol(1) + 1-nitropropane(2) ($T$ = 298.15 K); (2), 1-propanol(1) + nitromethane(2) ($T$ = 333.15 K).

**SUPPLEMENTARY MATERIAL**

# THERMODYNAMICS OF MIXTURES CONTAINING A VERY STRONGLY POLAR COMPOUND. 12. SYSTEMS WITH NITROBENZENE OR 1-NITROALKANE AND HYDROCARBONS OR 1-ALKANOLS

JUAN ANTONIO GONZÁLEZ[(1)*], FERNANDO HEVIA[(1)], LUIS FELIPE SANZ[(1)], ISAÍAS GARCÍA DE LA FUENTE[(1)] AND CRISTINA ALONSO-TRISTÁN[(2)]

[(1)]G.E.T.E.F., Departamento de Física Aplicada, Facultad de Ciencias, Universidad de Valladolid, Paseo de Belén, 7, 47011 Valladolid, Spain,

*e-mail: jagl@termo.uva.es; Fax: +34-983-423136; Tel: +34-983-423757

[(2)] Dpto. Ingeniería Electromecánica. Escuela Politécnica Superior. Avda. Cantabria s/n. 09006 Burgos, (Spain)

TABLE S1

Physical constants and reduction parameters[a] in the ERAS model for pure nitroalkanes at 298.15 K.

| Nitroalkane | $V_m$ / cm$^3$·mol$^{-1}$ | $10^3 \alpha_p$ /K$^{-1}$ | $10^{12} \kappa_T$ /Pa$^{-1}$ | $V_m^*$/ cm$^3$·mol$^{-1}$ | $P^*$ /J·cm$^{-3}$ |
|---|---|---|---|---|---|
| Nitromethane | 53.96 [s1] | 1.24 [s1] | 738[b] [s2,s3] | 41.67 | 840 |
| Nitroethane | 71.86 [s1] | 1.17 [s3] | 805[c] [s4,s5] | 56.09 | 711.7 |
| Nitrobenzene | 102.74 [s1] | 0.833 [s1] | 424[d] [s6,s7] | 84.73 | 861 |

[a] $V_m$, molar volume; $\alpha_p$ isobaric expansion coefficient; $\kappa_T$, isothermal compressibility; $V_m^*$, reduction parameters for molar volume; $P^*$, reduction parameter for pressure; [b] value determined using adiabatic compressibility (506.4 ·$10^{-12}$ Pa$^{-1}$) from [s2]; and isobaric molar heat capacity (106.62 J·mol$^{-1}$·K$^{-1}$ from [s3]; [c] value determined using adiabatic compressibility (586 ·$10^{-12}$ Pa$^{-1}$) from [s4]; and isobaric molar heat capacity (134.12 J·mol$^{-1}$·K$^{-1}$ from [s5]; [c] value determined using adiabatic compressibility (307·$10^{-12}$ Pa$^{-1}$) from [s6]; and isobaric molar heat capacity (181.13 J·mol$^{-1}$·K$^{-1}$ from [s7]

TABLE S2

Excess molar volumes, $V_m^E$, for 1-alkanol(1) + nitromethane(2), or + nitrobenzene(2) mixtures at equimolar composition and temperature $T$.

| System | $T$/K | $V_m^E$ / cm$^3$·mol$^{-1}$ | | Ref. |
|---|---|---|---|---|
| | | Exp[a] | ERAS[b] | |
| Methanol + nitromethane | 298.15 | − 0.164 | − 0.166 | S8 |
| Ethanol + nitromethane | 298.15 | 0.026 | 0.023 | S8 |
| 1-propanol + nitromethane | 298.15 | 0.235 | 0.253 | S8 |
| 1-butanol + nitromethane | 298.15 | 0.333 | 0.327 | S8 |
| Methanol + nitrobenzene | 298.15 | − 0.844 | − 0.580 | S9 |
| ethanol + nitrobenzene | 298.15 | − 0.671 | − 0.403 | S9 |
| 1-propanol + nitrobenzene[c] | 303.15 | − 0.1915 | − 0.188 | S10 |
| 1-butanol + nitrobenzene[c] | 303.15 | − 0.1172 | − 0.112 | S10 |

[a]experimental result; [b]ERAS calculations with parameters listed in Table 3; [c]for these systems results were obtained using $K_{AB}$ = 54.3.

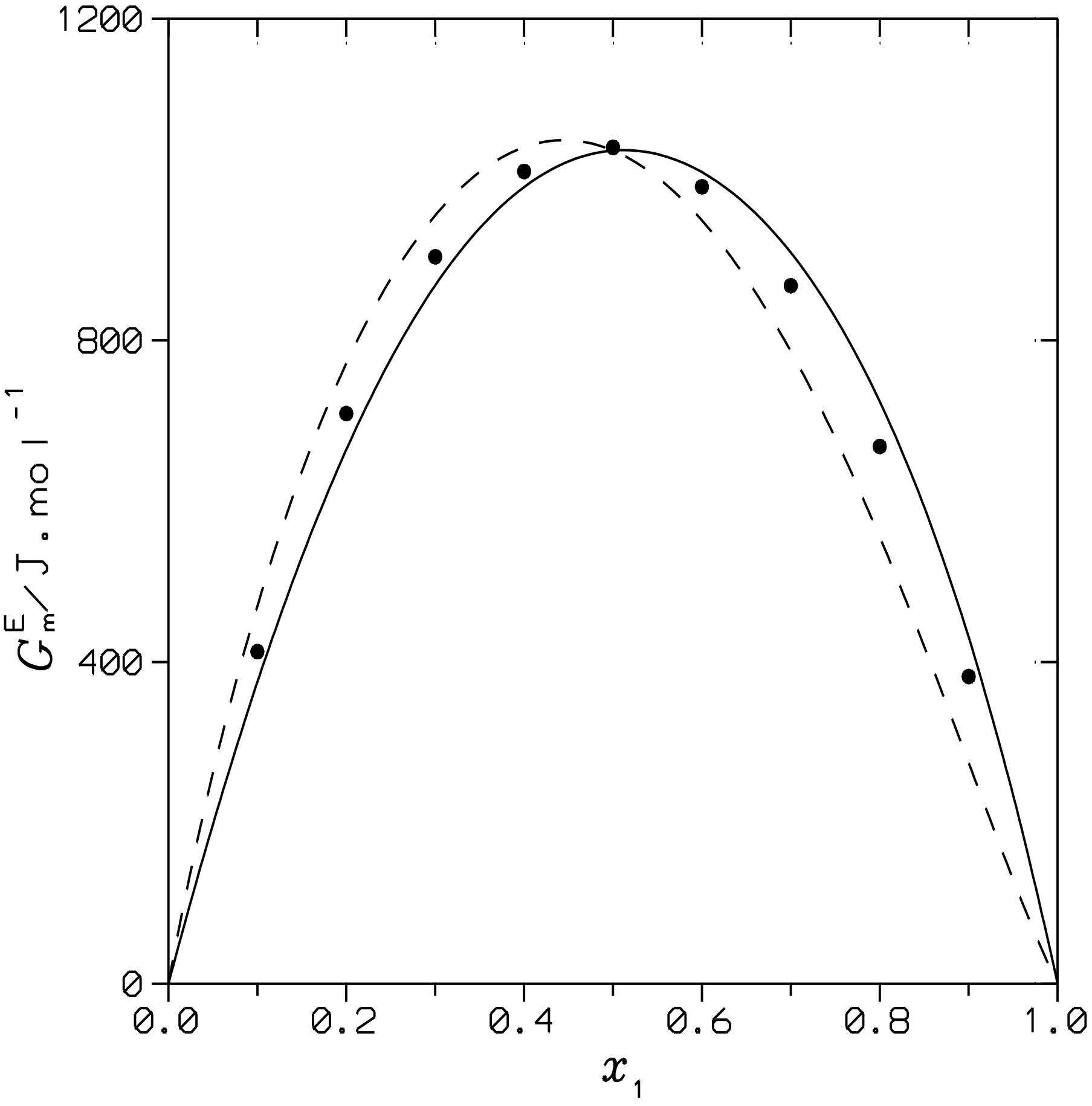

Figure S1 $G_m^E$ for the methanol(1) + nitromethane(2) mixture at 298.15 K Points, experimental results [s11]. Solid line, DISQUAC calculations. Dashed line, ERAS result using parameters from Table 3 and $Q_{AB}$ (entropy parameter needed for VLE calculations) = − 0.22 J·cm$^{-3}$·K$^{-1}$.

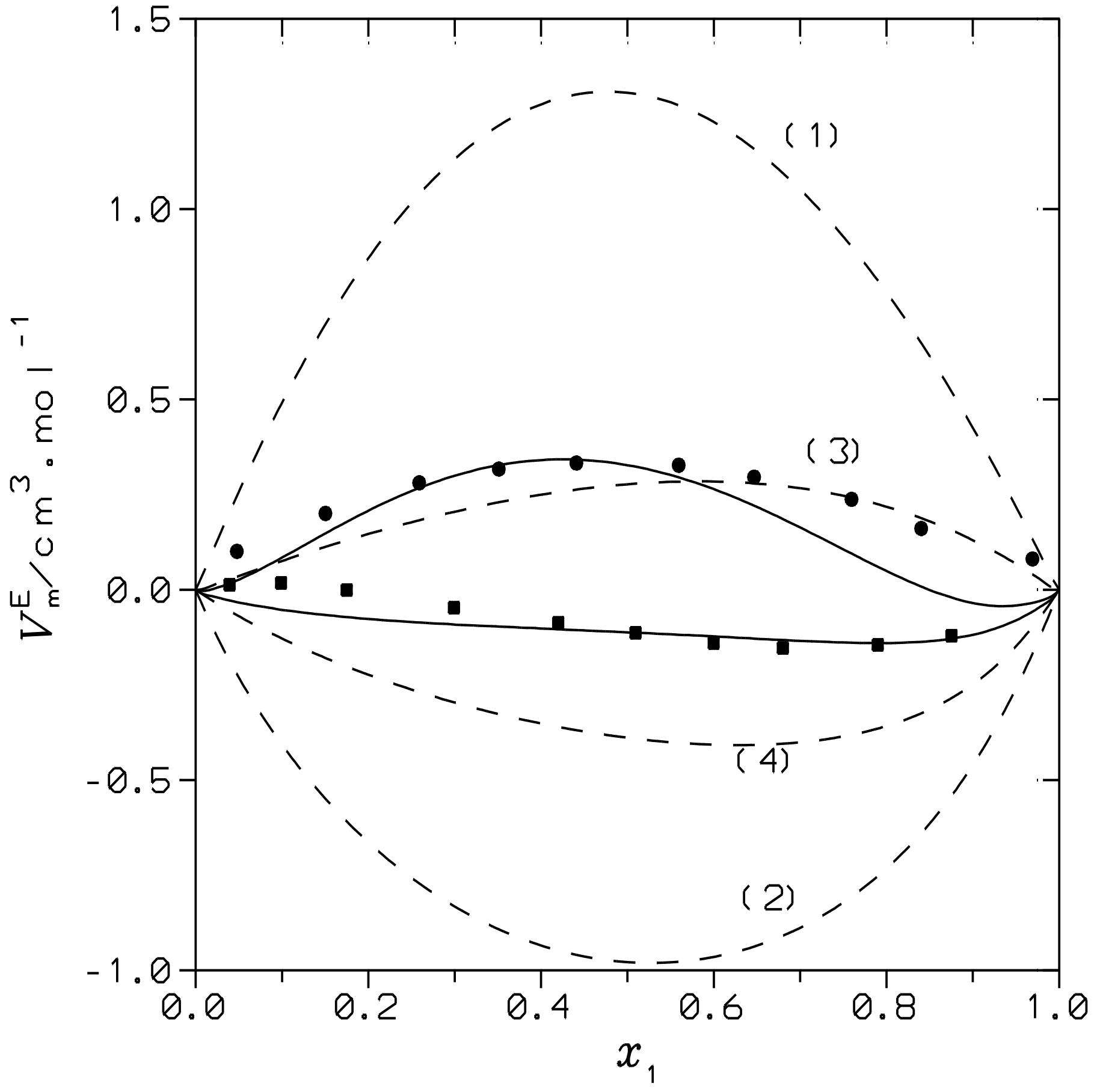

Figure S2 $V_m^E$ for the mixtures 1-butanol(1) + nitromethane(2) (●, [19]; $T$ = 298.15 K), or + nitrobenzene(2) (■, [s10]; $T$ = 303.15 K). Solid lines, ERAS results with parameters given in Table 3. Dashed lines, physical and chemical contributions in the ERAS model: (1) and (2), $V_{m,phys}^E$ and $V_{m,chem}^E$, respectively for the nitromethane solution; (3) and (4), $V_{m,phys}^E$ and $V_{m,chem}^E$ for the nitrobenzene mixture.

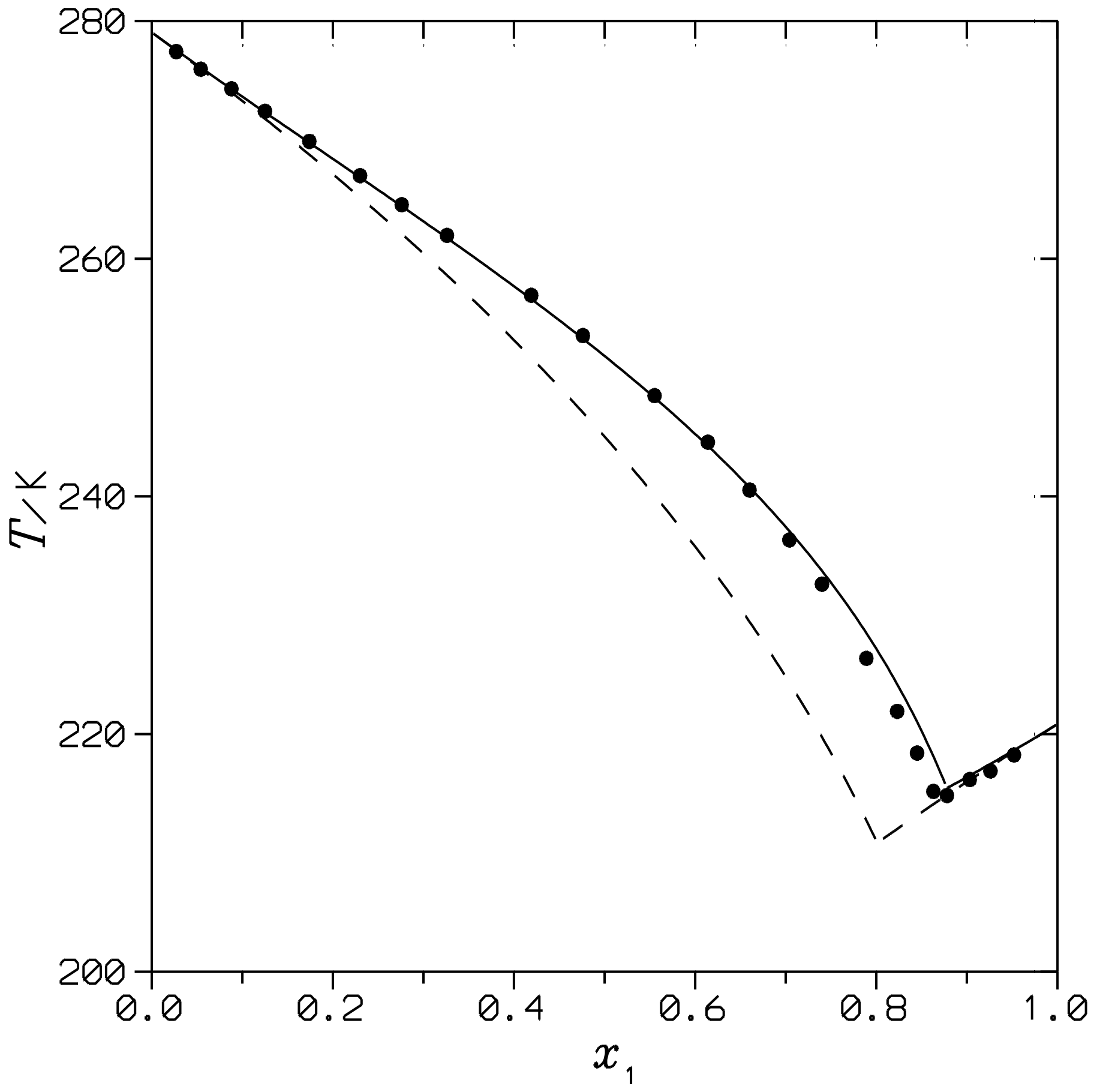

Figure S3 SLE for the nitrobenzene(1) + indane/2) mixture. Points, experimental values [s12], Solid lines, DISQUAC calculations (see Text). Dashed lines, results from the Ideal Solubility Model